\documentclass[12pt,a4paper]{article}
\pdfoutput=1

\usepackage{amssymb}
\usepackage{amsmath}
\usepackage{amsfonts}
\usepackage{amsthm}
\usepackage{mathrsfs}  				
\usepackage{enumitem}
\usepackage{setspace}
\usepackage{color}
\usepackage{bm}
\usepackage[pdftex]{graphicx}
\usepackage{authblk}
\usepackage{subfigure}
\usepackage{float}
\usepackage[utf8]{inputenc}
\usepackage{subfigure}  			
\usepackage[font=small]{caption}	
\usepackage[compress]{cite}         
\usepackage{float}

\usepackage{datetime}

\usepackage{pgfplots}   			
	\pgfplotsset{width=10cm}
	\usepackage{bbm}
	\usepackage{tensor}
	\usepackage{physics}
	\usepackage{slashed}

\numberwithin{equation}{section}

\newcommand{\BbbR}{\mathbb{R}}
\newcommand{\BbbZ}{\mathbb{Z}}

\DeclareMathOperator{\Realpart}{Re}

\theoremstyle{plain}

\usepackage{color}

\newdateformat{daymonthyear}{\THEDAY \, \monthname[\THEMONTH] \THEYEAR}
\newdateformat{monthyear}{\monthname[\THEMONTH] \THEYEAR}

\pgfplotsset{compat=1.17}

\begin{document}

\title{Vacua in locally
de Sitter cosmologies, 
\\
and how to distinguish them}
\author{Vladimir Toussaint$^{1}$\thanks{{\tt vladimir.toussaint@nottingham.edu.cn}} }
\author{Jorma Louko$^{2}$\thanks{\tt jorma.louko@nottingham.ac.uk}}
\affil{$^{1}$School of Mathematical Sciences,\\ 
University of Nottingham Ningbo China,\\
Ningbo 315100, PR China}
\affil{$^{2}$School of Mathematical Sciences, University of Nottingham, Nottingham NG7 2RD, UK}

\date{April 2023, revised August 2023\footnote{Published in Phys.\ Rev.\ D \textbf{109}, 025007 (2024), 
doi:10.1103/PhysRevD.109.025007. 
For Open Access purposes, 
this Author Accepted Manuscript is made available under CC BY public copyright.}}

\maketitle

\begin{abstract}
$(1+1)$-dimensional locally de Sitter Friedmann-Robertson-Walker cosmologies with compact spatial sections 
allow cosh, sinh and exponential evolution laws, 
each with a freely-specifiable spatial circumference parameter, 
and the value of this parameter has an invariant geometric meaning for the cosh and sinh evolution laws. 
We identify geometrically preferred states for a quantised massive scalar field
on these cosmologies, some singled out by adiabatic criteria in the distant past, 
with an ambiguity remaining due to a massive zero mode, 
and others induced from the Euclidean vacuum on standard $(1+1)$-dimensional de Sitter space by a quotient construction. 
We show that a comoving quantum observer, modelled as an Unruh-DeWitt detector, 
can distinguish these states from the Euclidean vacuum on standard de Sitter space. 
Numerical plots are given in selected parameter regimes. 
We also evaluate the field's stress-energy tensor expectation value 
for those states that are induced from the Euclidean vacuum by a quotient construction. 
\end{abstract}

\singlespacing


\newpage

\section{Introduction\label{sec:intro}}

$(d+1)$-dimensional de Sitter spacetime, $dS_{d+1}$, admits for all $d\ge1$ foliations as a spatially homogeneous Friedmann-Robertson-Walker (FRW) cosmology where the scale factor as a function of the cosmological time is a hyperbolic cosine, a hyperbolic sine, or an exponential. The case of four spacetime dimensions, $d=3$, is described in 
\cite{hawking-ellis}, and the generalisation to any $d\ge1$ is straightforward. 
For the cosh scale factor the foliation is global on~$dS_{d+1}$, 
whereas for the sinh and exp scale factors the foliation covers only part of~$dS_{d+1}$. 
For a quantised massive scalar field, the Euclidean vacuum in $dS_{d+1}$ 
\cite{chernikov-tagirov,tagirov,Bunch:1978yq,Birrell:1982ix,Allen:1985ux} hence induces on each of these cosmologies a quantum state, and to any local quantum observer this state is indistinguishable from the Euclidean vacuum on $dS_{d+1}$. 

In this paper we address how the situation changes when these FRW foliations of de Sitter are generalised to FRW cosmologies that are still locally de Sitter but have a compact spatial topology, even for the sinh and exp scale factors. 
Are there geometrically distinguished quantum states, and can a local quantum observer, modelled as an Unruh-DeWitt (UDW) detector~\cite{Unruh:1976db,DeWitt:1980hx,Birrell:1982ix}, operationally distinguish them from the Euclidean vacuum on de~Sitter? In $3+1$ dimensions, spatial quotients of this type have sometimes been explored as a possible description of our own Universe~\cite{ellis-schreiber,Levin:2001fg}. 

We specialise to $1+1$ spacetime dimensions, for two reasons. First, 
from the absence of spatial intrinsic curvature in $1+1$ dimensions it follows that the spatially compact locally de Sitter FRW spacetimes can be classified by a spatial circumference parameter that can take arbitrary positive values, 
for each of the cosh, sinh and exp scale factors; 
for the exp scale factor the spatial circumference parameter has no geometrically invariant magnitude, 
but for the cosh and sinh scale factors it does, as we shall review in Section~\ref{sec:De-Sitter DS}. 
We emphasise that the only one of these spacetimes that is a subset of $dS_{1+1}$ is the global foliation of $dS_{1+1}$, given by the cosh scale factor with the standard spatial circumference: 
none of the others is globally isometric to $dS_{1+1}$ or to any of its subsets. 

The second reason to specialise to $1+1$ dimensions is that a pointlike UDW detector operating for a finite time at constant coupling encounters then no singularities in its response, to leading order in perturbation theory, as we shall review in Section~\ref{sec:comoving-detectors}. 
This will simplify the technical input needed to localise the detector in time. 

Our main outcome is to present a selection of geometrically distinguished sets of quantum states for these locally de Sitter cosmologies, 
and to show that a comoving observer can use their local quantum equipment to establish that these states differ from the Euclidean vacuum in $dS_{1+1}$. 

For the sinh and exp scale factor spacetimes, we consider the vacuum that is adiabatic in the past, which is a criterion often adopted in the $(3+1)$ context for describing our own Universe~\cite{Mukhanov:2007zz}; 
for the exp scale factor spacetime without spatial periodicity, this criterion would in fact give the standard Euclidean vacuum~\cite{Bunch:1978yq,Birrell:1982ix}. 
The distant past adiabatic criterion is however inapplicable to the spatially constant field mode that appears due to the spatial compactness, because of a phenomenon known as a massive zero mode~\cite{Ford:1989mf,Toussaint:2021czo}. 
We describe a possible alternative way to choose the quantum state 
of the spatially constant field mode by appealing to the late-time behaviour. 

For the cosh scale factor spacetimes, we consider the adiabatic large momentum vacuum that mimics the Euclidean vacuum, and is in fact induced from the Euclidean vacuum by a quotient construction when spatial circumference parameter takes those special values for which the spacetime is a $\BbbZ_p$ quotient of standard de~Sitter, with $p=2,3\ldots$. 

For the sinh scale factor spacetimes, we also consider, 
in addition to the adiabatic vacuum, 
a state induced from the Euclidean vacuum by a quotient construction, 
allowing a straightforward method-of-images construction of the Wightman function. 
When described in the Fock space formalism adapted to the cosmological evolution, 
this state is however not pure but mixed, 
much like the Minkowski vacuum in a single Rindler wedge appears 
as as a thermal state with respect to the Rindler time evolution~\cite{Unruh:1976db,Birrell:1982ix}. 

In all of these situations, we find the response of a comoving UDW detector, 
either as a mode sum or as a double integral involving a hypergeometric function. 
We present selected numerical results. 

For comparison, we also evaluate the scalar field's stress-energy tensor in those states that come from the Euclidean vacuum by a quotient construction, in terms of image sums involving a hypergeometric function. 
These image sums provide a starting point for surveying the parameter space for the stress-energy tensor 
by analytic asymptotic methods and by numerical methods. 

The structure of the paper is as follows. 
We start in Section 
\ref{CompactifiedCosmol}
with a brief review of a quantised real massive scalar field in $(1+1)$-dimensional 
spatially compact FRW spacetimes, establishing the notation and the conventions. 
Section 
\ref{sec:De-Sitter DS}
describes the geometry of the three distinct types of locally de~Sitter $1+1$ spacetimes that 
are FRW cosmologies with compact spatial sections. 
Section 
\ref{sec:QFT-setup} 
presents our choices for the state of the quantum field, 
and the formulas for the response of a comoving UDW detector are given in 
Section~\ref{sec:comoving-detectors}. 
The image sum formulas for the stress-energy tensor are given in 
Section~\ref{sec:stress-energy}. 
Section \ref{sec:numerics} describes the outcomes of selected numerical simulations, 
delegating the associated plots to an appendix. 
Section \ref{sec:conclusions} gives a brief summary and concluding remarks. 

We use units in which $\hbar = c =1$. 
The geometry conventions follow those in~\cite{Birrell:1982ix}, 
having $ds^2>0$ for timelike separations. 
Complex conjugation is denoted by an asterisk,
except in selected long expressions where it is denoted by an overline.

\section{Massive scalar field in spatially compact $(1+1)$
FRW spacetimes\label{CompactifiedCosmol}} 

In this section we briefly review the quantisation of a real massive scalar field in $(1+1)$-dimensional 
spatially compact FRW spacetimes, 
setting out the notation and the conventions, which follow those in~\cite{Toussaint:2021czo}. 

\subsection{Spacetime and the field equation}

We consider spacetimes with the line element 
\begin{align}\label{ExpComosMetric}
ds^2 = dt^2 - a^2(t)dx^2 = C(\eta)(d\eta^2 - dx^2) \, , 
\end{align}
where the scale factor $a(t)$ is assumed positive, 
the cosmological time $t$ and the conformal time $\eta$ are related by $d\eta/dt = 1/a(t)$, 
and $C(\eta) = a^2 \bigl(t(\eta)\bigr)$. 
$C$~is by assumption positive, and we assume it to be smooth in~$\eta$. 
We further assume $x$ to be periodic with period $L>0$, 
so that $(t, x)\sim(t, x+L)$, or $(\eta, x)\sim(\eta, x+L)$. 
The constant $\eta$ surfaces are hence topologically circles, and the 
spacetime has topology $\BbbR \times S^1$. 

We take $t$ and $a$ to have the physical dimension of length. It follows that 
$\eta$ and $x$ are dimensionless and 
$C$ has the physical dimension of length squared. This convention will be useful with the locally 
de Sitter metrics introduced in Section \ref{sec:De-Sitter DS}. 

On these spacetimes, we consider a real scalar field $\phi$ with mass $m>0$ and the action 
\begin{align}
S = \frac{1}{2}  \int \left [( \partial_\eta \phi)^2 -(\partial_x \phi)^2 - \tilde{\mu}^2 (\eta) \phi^2\right] d\eta\,dx 
\, , 
\label{eq:action}
\end{align}
where
\begin{align}
\tilde{\mu}(\eta) :=\bigl(C(\eta)(m^2 + \xi R)\bigr)^{1/2}
\, , 
\label{eq:mutilde-def}
\end{align}
$R$ is the Ricci scalar, $\xi$ is the curvature coupling constant, 
and we assume the parameters to be such that $\tilde{\mu}$ is positive. 
The field equation is the Klein-Gordon equation, 
\begin{align}\label{scalar_fieldEqI}
\left(\frac{\partial^2}{\partial \eta^2}  - \frac{\partial^2}{\partial x^2} + \tilde{\mu}^2(\eta)\right) \phi(\eta, x) = 0 \, , 
\end{align}
where $\tilde{\mu}$ acts as an $\eta$-dependent effective mass. 

We assume $\phi$ to be a single-valued field on the spacetime, 
periodic in the local coordinate $x$ with period~$L$. 
(A generalisation to a $\BbbZ_2$-twisted field is considered in~\cite{Toussaint:2021czo}.) 
We may hence separate the field equation \eqref{scalar_fieldEqI} with the normal mode ansatz 
\begin{align}\label{Scalar field: discrete normal modes}
U_n(\eta,x) =  L^{-1/2}\chi_n(\eta)\exp( i k_n x)
\, ,
\end{align}
where 
\begin{align}\label{eq:cylindermomenta_ScalarField_FRW}
k_n &:= 
2\pi n /L\, , \quad n\in\BbbZ\, ,
\end{align}
finding that mode function $\chi_n$ satisfies the differential equation
\begin{align}
\label{ScalarFrequency_ode-discrete}
\chi''_n(\eta) + \omega_n^2(\eta)\chi_n(\eta) \,= 0 \, , 
\end{align}
where the prime denotes derivative with respect to $\eta$ and 
\begin{align}
\label{ScalarFrequency_def-discrete}
\omega_n(\eta) &:= \bigl(k_n^2 +  \tilde{\mu}^2(\eta)  \bigr)^{1/2} \, 
\, . 
\end{align}
We require the mode functions to satisfy the Wronskian condition 
\begin{align}\label{WronskianNorma:CompactifiedSpa}
W[\chi_n,\chi_n^*] := \chi_n  \chi^{\prime *}_n - \chi_n^*  \chi'_n = i
\, , 
\end{align}
which implies that the normal modes 
$U_n$ \eqref{Scalar field: discrete normal modes}
are a positive norm orthonormal set in the Klein-Gordon inner product, 
\begin{align}\label{Hilbert_scalarproduct:CompactifiedSpa}
( U_n, U_{n'}) := i  \int_{0}^{L} d x  \, 
\left( U_n^*\partial_\eta U_{n'} - U_{n'}\partial_\eta U_n^* \right) =\delta_{n{n'}} 
\, . 
\end{align}

\subsection{Field quantisation\label{subsec:untwisted-theory}}

Given a choice of the mode functions~$\chi_n$, 
we quantise the field in a standard manner, expanding the field operator 
$\hat\phi$ as 
\begin{align}
\hat{\phi}(\eta, x) 
= \sum_n 
\Bigl(
U_{n}(\eta, x)  \hat{a}_n
+  h.c.
\Bigr) 
\, ,
\label{eq:psi-osc+zm-decomposition}
\end{align}
where the 
nonvanishing commutators of the annihilation and creation operators are
\begin{align}
\label{commutatoRelat_ExpanCosmII}
\bigl[ \hat{a}_n, \hat{a}^\dag_m\bigr] 
= \delta_{n,m} \mathbbm{1} 
\ . 
\end{align}
$\hat{\phi}(\eta,x)$ and its conjugate momentum 
$\hat{\pi}(\eta,x) = \frac{\partial}{\partial\eta} \hat{\phi}(\eta,x)$
then satisfy the canonical commutation relations, 
\begin{align}
\left[ \hat{\phi}(\eta,x), \hat{\pi}(\eta, x') \right] 
=i \delta_{x,x'} 
\mathbbm{1} 
\, ,
\label{eq:hatphi-hatpi-commutator}
\end{align}
where $\delta_{x,x'}$ is the periodic Dirac delta-function. 

The Hilbert space is a Fock space built on the Fock vacuum $|0\rangle$, 
satisfying $a_n|0\rangle =0$ for all~$n$. 
The Wightman function, $G(\eta,x ; \eta', x') = \langle0|\hat{\phi}(\eta,x) \hat{\phi}(\eta',x') |0\rangle$, decomposes as 
\begin{subequations}
\begin{align}
G(\eta,x ; \eta', x') 
&= 
G_0(\eta; \eta') 
+ 
G_{osc}(\eta,x ; \eta', x') 
\, , 
\\
G_0(\eta; \eta') 
&= 
\frac{1}{L} \chi_0(\eta)\chi_0^*(\eta')
\, , 
\label{eq:Gnought-pre}
\\
G_{osc}(\eta,x ; \eta', x') 
&= 
\frac{1}{L} \sum_{n\ne0}  \chi_n (\eta)\chi_n^*(\eta') e^{ik_n(x - x')} 
\, . 
\label{eq:Gosc-pre}
\end{align}
\label{eq:G-totalsum}
\end{subequations}

The mode functions $\chi_n$ can often be chosen by requiring them 
to have the 
adiabatic asymptotic form \cite{Birrell:1982ix,Mukhanov:2007zz,Parker:1974qw,Audretsch:1979uv}, 
\begin{align}
\label{pos:frequency_ScalarMode_nonstatic}
\chi_n(\eta)  
\longrightarrow  
\frac{1}{ \sqrt{2\omega_n(\eta)} }\exp(-i\int^{\eta}\omega_n(\eta')d\eta') \, ,
\end{align}
where $\omega_n(\eta)$ is slowly varying in a suitable technical sense. 
The sense of asymptotic may refer to a future/past region, $\eta\to\pm\infty$, 
or to large momenta, 
$|n|\to\infty$.  
The vacuum state that ensues is interpreted physically as 
a no-particle state for a local observer, 
respectively in the asymptotic region, or for the asymptotically large momenta. 

We shall encounter spacetimes with an asymptotic past region, $\eta\to-\infty$, 
in which $C(\eta) \to 0$ in such a way that mode functions with the adiabatic form \eqref{pos:frequency_ScalarMode_nonstatic} exist 
for $n\ne0$ but not for $n=0$. 
An $n=0$ mode with this property is 
known as a massive zero mode~\cite{Ford:1989mf}.

\section{Locally de Sitter FRW spacetimes\label{sec:De-Sitter DS}}

In this section we review the geometry of $(1+1)$-dimensional de~Sitter spacetime, 
adapting the $(3+1)$-dimensional discussion of~\cite{hawking-ellis}, 
and introducing coordinate charts whose generalisation will then 
lead to three distinct families of spatially compact locally but not globally de Sitter FRW spacetimes. 

\subsection{de Sitter spacetime}

$(1+1)$-dimensional de~Sitter spacetime $dS_2$ can be defined as the hyperboloid 
\begin{align}
-H^{-2} = T^2 - X^2 - Y^2 \, ,
\label{eq:deS-hyperboloid}
\end{align}
embedded in $(2+1)$-dimensional Minkowski spacetime with global coordinates $(T, X, Y)$ and the metric
\begin{align}
ds^2 = dT^2 - dX^2 -dY^2 \, , 
\label{eq:ambient-Minkowski-metric}
\end{align}
where $H$ is a positive parameter of dimension inverse length.  
$dS_2$ is a maximally symmetric spacetime, with the isometry group $O(2,1)$ inherited from the ambient Minkowski space, and with the positive Ricci scalar $R = 2H^2$, showing that $H$ is the inverse of the Gaussian curvature radius. 
$dS_2$ is globally hyperbolic, with spatial topology~$S^1$, 
and a global $\BbbR \times S^1$ foliation 
is provided by the spacelike circles of constant~$T$. 

For later convenience, we define \cite{Allen:1987tz,Allen:1985ux}
\begin{align}\label{RealQuadraticForm}
\mathcal{Z}(\mathsf{x},\mathsf{x}') := - H^2\eta_{ab}X^a(\mathsf{x})X^b(\mathsf{x}') \, ,
\end{align}
where $\mathsf{x}$ and $\mathsf{x}'$ denote points in~$dS_2$, 
$X^a(\mathsf{x})$ and $X^b(\mathsf{x}')$ are the corresponding three-dimensional Minkowski coordinates, 
and $\eta_{ab}$ is the ambient Minkowski metric~\eqref{eq:ambient-Minkowski-metric}. 
This real quadratic form is clearly invariant under the de~Sitter group $O(2,1)$, 
and it has the properties that $\mathcal{Z}(\mathsf{x},\mathsf{x}') > 1$ if $(\mathsf{x},\mathsf{x}')$ are timelike separated, $\mathcal{Z}(\mathsf{x},\mathsf{x}')= 1$ if $(\mathsf{x},\mathsf{x}')$ are null separated and $\mathcal{Z}(\mathsf{x},\mathsf{x}') < 1$ if $(\mathsf{x},\mathsf{x}')$ are spacelike separated. 
Further discussion is given in~\cite{Allen:1987tz, Allen:1985ux}. 

We now consider three different coordinate charts, 
adapted to different subgroups of the $O(2,1)$ isometry group, 
and use them to construct three distinct families of 
spatially compact locally but not globally de Sitter FRW spacetimes. 

\subsection{Cosh scale factor spacetime}

The first chart of interest on $dS_2$ is $(t,\chi)$, in which 
\begin{align}\label{closedSS}
T = H^{-1} \sinh(Ht),  \quad     X = H^{-1}\cosh(Ht)\cos \chi, \quad Y = H^{-1}\cosh(Ht) \sin \chi \, ,
\end{align}
where $-\infty <t< \infty$ and 
$(t,\chi) \sim (t, \chi + 2\pi)$. 
This chart covers all of $dS_2$, and is often called the global chart. 
The metric reads 
\begin{align}
\label{ClosedSpatialSection-dS}
ds^2 = dt^2 - H^{-2}\cosh^2(Ht)d\chi^2 \, .
\end{align}
Introducing the conformal time $\eta$ as in~\eqref{ExpComosMetric}, we have 
\begin{align}
\eta := 2 \arctan\left(e^{Ht}\right) \, , 
\end{align}
where $0 < \eta < \pi$, and 
\begin{align}
ds^2 = (H \sin \eta)^{-2}(d\eta^2 - d\chi^2)\, .
\label{eq:coshchart-metric}
\end{align}
The spacelike circles of constant $\eta$ are the circles of 
constant $T$ in~\eqref{eq:deS-hyperboloid}. 

We shall consider the metric \eqref{eq:coshchart-metric} 
where $\chi$ is periodic with period $\lambda\in(0,\infty)$, 
not necessarily equal to $2\pi$. 
Geometrically, this means first passing to the universal covering space of $dS_2$ 
and then taking the quotient by the discrete group generated by the spatial translation $(\eta,\chi) \mapsto (\eta, \chi + \lambda)$, coming from the rotational Killing vector $X\partial_Y - Y\partial_X$ of the embedding spacetime. 
We refer to this spacetime as the cosh scale factor spacetime. 

We shall be especially interested in the cases where $\lambda = 2\pi/p$, $p = 1,2,\ldots$, 
where $p=1$ is normal $dS_2$ and $p = 2,3,\ldots$ are its $Z_p$ quotients; 
the quotient with $p=2$ was preferred by de Sitter in the $(3+1)$-dimensional context~\cite{desitter-mnras,McInnes:2003xm}. 
For the case $\lambda = 4\pi$, which is the double cover of normal~$dS_2$, 
a real scalar field has been considered in \cite{Epstein:2020wgf}, 
by the techniques developed for normal $dS_2$ in \cite{Bros:1994dn,Bros:1995js,Bros:1998ik}, 
and a complex automorphic field has been considered in~\cite{Higuchi:2022nfy}.

\subsection{Sinh scale factor spacetime}

A second chart of interest on $dS_2$ is $(t_1,\chi_1)$, in which 
\begin{align}
T = H^{-1} \sinh(Ht_1)\cosh\chi_1,  \quad     X = H^{-1} \sinh(Ht_1)\sinh\chi_1, \quad Y = H^{-1} \cosh(Ht_1) \, ,
\label{eq:sinhchart-embedding}
\end{align}
where $ 0 < t_1< \infty$ and $\chi_1 \in \BbbR$. 
This chart covers the part of $dS_2$ where $Y > 1/H$ and 
$T = \sqrt{X^2 + \left( Y^2 - {H}^{-2}\right)}$, so that in particular $T>|X|$. 
The metric reads 
\begin{align}
\label{OpenSpatialSection-DS}
ds^2 = dt_1^2 - H^{-2}\sinh^2(Ht_1)d\chi_1^2 \, .
\end{align}
Introducing the conformal time $\eta_1$ as in~\eqref{ExpComosMetric}, we have 
\begin{align}
\eta_1 := \ln\bigl(\tanh(Ht_1/2)\bigr) \,, 
\end{align}
where $-\infty < \eta_1 < 0$, 
and 
\begin{align}
ds^2 = (H \sinh \eta_1)^{-2}(d\eta_1^2 - d\chi_1^2)\, .
\label{eq:sinhchart-metric}
\end{align}
The spacelike curves of constant $\eta_1$ are the hyperbolas
$T = \sqrt{X^2 + \left( Y^2 - {H}^{-2}\right)}$ 
in the timelike planes of constant $Y>1/H$. 

The coordinates $(t_1,\chi_1)$ are similar to the Milne coordinates in the future quadrant of $(1+1)$ Minkowski spacetime, and the coordinate singularity at $t_1 \to 0^+$, or $\eta_1 \to -\infty$, is similar to the singularity of the Milne coordinates on the Minkowski light cone 
\cite{Birrell:1982ix,Toussaint:2021czo}. 

We shall consider the metric \eqref{eq:sinhchart-metric} 
where $\chi_1$ is periodic with period $\lambda_1\in(0,\infty)$. 
Geometrically, this means taking the quotient by the group generated by the translation 
$(\eta_1,\chi_1) \mapsto (\eta_1, \chi_1 + \lambda_1)$, 
coming from the boost Killing vector $X\partial_T + T\partial_X$ of the embedding spacetime. 
We refer to this spacetime as the sinh scale factor spacetime.

\subsection{Exp scale factor spacetime}

A third chart of interest on $dS_2$ is $(t_0,\chi_0)$, in which 
\begin{subequations}
\begin{align}
T &= H^{-1}\sinh(Ht_0) + \tfrac{1}{2}H^{-1}\chi^2_0\exp(Ht_0) \,, \\
X &= H^{-1}\chi_0 \exp(Ht_0) \,,\\
Y &= H^{-1}\cosh(Ht_0) - \tfrac{1}{2}H^{-1}\chi^2_0\exp(Ht_0)\,,
\end{align}
\end{subequations} 
where $-\infty <t_0< \infty$ and $\chi_0 \in\BbbR$. 
This chart covers the part of $dS_2$ where $T+Y > 0$. 
The metric reads 
\begin{align}
ds^2 = dt^2_0 - H^{-2}e^{2Ht_0} d\chi^2_0\, .
\label{eq:expchart-metric-cosmoltime}
\end{align}
Introducing the conformal time $\eta_0$ as in~\eqref{ExpComosMetric}, we have 
\begin{align}
\eta_0:=     -e^{-Ht_0}\, ,
\end{align}
where $-\infty < \eta_0 < 0$, and 
\begin{align}
ds^2 = (H \eta_0)^{-2}(d\eta_0^2 - d\chi_0^2)\, .
\label{eq:expchart-metric}
\end{align}
The spacelike curves of constant $\eta_0$ are the parabolas
$T - Y = (X^2 - H^{-2})/(T+Y)$ 
in the null planes of constant $T+Y>0$. 

The coordinates $(t_0,\chi_0)$ have a coordinate singularity at $t_0 \to -\infty$, 
or $\eta_0 \to -\infty$, at $T+Y=0$.
The higher-dimensional version of these coordinates are known as the 
spatially flat coordinates in de~Sitter~\cite{hawking-ellis}, 
much employed in cosmology~\cite{Mukhanov:2007zz}. 

We shall consider the metric \eqref{eq:expchart-metric} 
where $\chi_0$ is periodic with period $\lambda_0\in(0,\infty)$. 
Geometrically, this means taking the quotient by the group generated by the translation 
$(\eta_0,\chi_0) \mapsto (\eta_0, \chi_0 + \lambda_0)$, 
coming from the null rotation Killing vector 
$(X\partial_T + T\partial_X) - (X\partial_Y - Y\partial_X)$ of the embedding spacetime. 
We refer to this spacetime as the exp scale factor spacetime. 

The value of the parameter $\lambda_0\in(0,\infty)$ has no geometrically invariant meaning: 
any other value $\tilde\lambda_0\in(0,\infty)$ gives an isometric spacetime, 
as can be seen from \eqref{eq:expchart-metric} by the coordinate transformation 
\begin{align}
(\eta_0,\chi_0) = ( k\tilde\eta_0, k \tilde\chi_0\bigr)
\ , 
\label{eq:expchart-rescaling}
\end{align}
where $k = \lambda_0/\tilde\lambda_0$. 
This redundancy comes about because a null rotation in the embedding Minkowski spacetime 
has no Lorentz-invariant magnitude, 
in contrast to the rotation angle of a rotation and the rapidity parameter of a boost. 
In particular, we could set $\lambda_0 = 1$ without loss of generality. 
We shall however keep $\lambda_0$ general as this will give insight in the choice of the spatially constant mode state in 
Sections \ref{sec:exp-chart-QFT} and~\ref{sec:exp-chart-detector}.

\section{Quantum field in locally de Sitter FRW spacetimes\label{sec:QFT-setup}}

In this section we consider the quantised scalar field in our three families of 
spatially compact locally de~Sitter FRW spacetimes. 
The key aim is to identify distinguished vacuum states 
and to write down their Wightman functions. 

We assume throughout $m^2 + 2 \xi H^2>0$, which makes 
the effective mass squared in the Klein-Gordon equation strictly positive, and will 
in particular avoid the special issues that arise 
with a massless minimally coupled scalar field \cite{Allen:1987tz,Allen:1985ux,Kirsten:1993ug}. 
We write 
\begin{align}
\label{HankelFunct_index}
\nu := \sqrt{\tfrac{1}{4} - m^2/H^2 - 2 \xi}\, ,
\end{align}
where we take the square root to be non-negative for $0 < m^2/H^2 + 2 \xi \le \tfrac14$, 
in which case $0 \le \nu< \tfrac12$, 
and positive imaginary for $\tfrac14 <  m^2/H^2 + 2 \xi$, 
in which case we write $\nu = i\alpha$ with $\alpha = \sqrt{m^2/H^2 + 2 \xi - \tfrac{1}{4}} > 0$.

\subsection{Cosh scale factor FRW\label{sec:coshchart-quantisation}}

For the cosh scale factor spacetime~\eqref{eq:coshchart-metric}, 
with $\chi$ periodic with period $\lambda \in (0,\infty)$, the formalism of 
Section \ref{CompactifiedCosmol} applies with $L = \lambda$ and $C(\eta) = (H\sin \eta)^{-2}$. 

A pair of linearly independent solutions to the mode function differential equation \eqref{ScalarFrequency_ode-discrete} are 
$\sqrt{\sin\eta} \, \mathsf{P}^\nu_{-\frac{1}{2}+ |k_n|}(-\cos\eta)$ 
and 
$\sqrt{\sin\eta} \, \mathsf{Q}^\nu_{-\frac{1}{2}+ |k_n|}(-\cos\eta)$, 
where $\mathsf{P}^\nu_{-\frac{1}{2}+ |k_n|}$ and $\mathsf{Q}^\nu_{-\frac{1}{2}+ |k_n|}$ are the 
associated Legendre functions on the cut, known in \cite{dlmf} as Ferrers functions. 
We choose the linear combinations 
\begin{align}
\label{ModeFunc-ClosedSPS}
\chi_n(\eta) = \sqrt{\sin\eta} \, 
e^{-i\pi\nu/2}
\left( \frac{\pi\Gamma(\frac{1}{2} -\nu+  |k_n|  )}{4\Gamma ( \frac{1}{2} +\nu + |k_n |)} \right)^\frac{1}{2}
\left(\mathsf{P}^\nu_{-\frac{1}{2}+ |k_n|}(-\cos\eta) -\frac{2i}{\pi} \mathsf{Q}^\nu_{-\frac{1}{2}+ |k_n|}(-\cos\eta)\right)\, ,
\end{align}
which satisfy the Wronskian condition \eqref{WronskianNorma:CompactifiedSpa} 
by the Wronskian properties of the Ferrers functions~\cite{dlmf}. 
The phase convention in \eqref{ModeFunc-ClosedSPS} covers the full range of $\nu$ by one formula: 
when $0 < \nu < \tfrac12$, the factor $e^{-i\pi\nu/2}$ 
could be replaced by unity, and when $\nu = i\alpha$ with $\alpha>0$, 
the ratio of the gamma-functions could be replaced by unity, as this would only affect the overall phase. 
We note in passing that the corresponding formula (5.74) in \cite{Birrell:1982ix} appears to contain a typographic error in the exponential factor. 

An alternative expression for $\chi_n$ \eqref{ModeFunc-ClosedSPS}, 
found via the connection formulas on p168 of~\cite{MOS1966}, is 
\begin{align}
\chi_n(\eta) 
&= 
e^{-i\pi/4 + i|k_n|\pi}
\frac{\sqrt{\Gamma\bigl(\tfrac12 + \nu + |k_n|\bigr)
\Gamma\bigl(\tfrac12 - \nu + |k_n|\bigr)}}{\sqrt{2} \, \Gamma(1 + |k_n|)}
\notag
\\
& \hspace{3ex}
\times e^{-i |k_n| \eta}
\, 
{}_2 F_1 
\! 
\left(
\tfrac{1}{2} + \nu, \tfrac{1}{2} - \nu;1 + |k_n|; 
\tfrac12 (1 + i \cot\eta) 
\right)\, ,
\label{ModeFunc-ClosedSPS-alt}
\end{align}
where ${}_2 F_1$ is the hypergeometric function~\cite{dlmf}. 
At large~$|n|$, \eqref{ModeFunc-ClosedSPS-alt} shows that $\chi_n(\eta)$ reduce to the standard plane waves, 
$\exp(-i|k_n|\eta)(2|k_n|)^{-1/2}$, 
and the state defined by these 
modes is in this sense an adiabatic vacuum at all times~\cite{Birrell:1982ix}. 
For $\lambda=2\pi$, in standard~$dS_2$, 
these modes define the state~$|0_E \rangle$, 
known as the Euclidean, Chernikov-Tagirov or Bunch-Davies vacuum 
\cite{chernikov-tagirov,tagirov,Bunch:1978yq,Birrell:1982ix,Allen:1985ux}. 

The spacetime is not slowly expanding at early or late times, 
in the terminology of Section~\ref{subsec:untwisted-theory}, 
and there is no adiabatic in region or out region. 
For later use, we record here that at late times, $\eta \to \pi^-$, 
the asymptotic behaviour of the modes 
\eqref{ModeFunc-ClosedSPS} can be found using the hypergeometric representations
in \S{}14.3 in~\cite{dlmf}: 
with a convenient phase choice, 
the leading two terms for $\nu\ne0$ are 
\begin{align}
\chi_n (\eta)
&\sim 
\frac{\sqrt\pi \, {(\pi-\eta)^{1/2}}}{2 \sin(\pi\nu) 
\sqrt{\Gamma\bigl(\tfrac12 + \nu + |k_n|\bigr) \Gamma\bigl(\tfrac12 - \nu + |k_n|\bigr)}}
\notag 
\\
& 
\hspace{3ex}
\times 
\left[
\frac{ e^{i\pi\nu/2} \, 2^\nu \Gamma\bigl(\tfrac12-\nu+|k_n|\bigr)}{\Gamma(1-\nu)} 
{(\pi-\eta)}^{-\nu}
- 
\frac{ e^{-i\pi\nu/2} \, 2^{-\nu} \Gamma\bigl(\tfrac12+\nu+|k_n|\bigr)}{\Gamma(1+\nu)} 
{(\pi-\eta)}^{\nu}
\right] 
\,,
\label{eq:Evac-modes-as-nugen}
\end{align}
and the expression for $\nu=0$ is 
\begin{align}
\chi_n (\eta)
\sim 
\frac{{(\pi-\eta)}^{1/2}}{\sqrt\pi}
\left[
\tfrac12 \pi i   + \ln2 - \gamma - \psi\bigl(\tfrac12 + |k_n|\bigr) - \ln(\pi-\eta)
\right] \,, 
\label{eq:Evac-modes-as-nu0}
\end{align}
where $\psi$ is the digamma function 
and $\gamma$ is the Euler-Mascheroni constant~\cite{dlmf}. 

From now on we consider only standard $dS_2$ 
and its $\BbbZ_p$ quotients $M_p$ in which $\lambda = 2\pi/p$, $p = 2,3,\ldots$. 
On~$M_p$, the Wightman function in the vacuum $|0_p\rangle$ 
defined by the modes \eqref{ModeFunc-ClosedSPS}
can then be obtained from the Euclidean vacuum $|0_E\rangle$ on $dS_2$ by the method of images \cite{Banach:1978dt,Banach:1979wz}, as follows. 

Recall first that the Euclidean vacuum Wightman function on $dS_2$ reads 
\cite{Bunch:1978yq} 
\begin{align}
\label{WightmanFunct_Standard_dS}
G_{dS_2}(\mathsf{x}, \mathsf{x}') 
= 
\tfrac{1}{4}\sec(\pi \nu) \, {}_2 F_1 
\Bigl(
\tfrac{1}{2} - \nu, \tfrac{1}{2} + \nu;1; 
\tfrac{1}{2} \bigl( 1+\mathcal{Z}_\epsilon(\mathsf{x}, \mathsf{x}') \bigr) 
\Bigr)\, ,
\end{align} 
where 
\begin{align}
\mathcal{Z}_\epsilon(\mathsf{x}, \mathsf{x}')
= 
\mathcal{Z}(\mathsf{x}, \mathsf{x}')
- i \epsilon H \bigl( {\tilde T}(\mathsf{x})-{\tilde T}(\mathsf{x}') \bigr)  - \epsilon^2 \,,
\label{eq:Z-standard-dS}
\end{align}
${\tilde T}$ is any global time function on~$dS_2$, 
and the limit $\epsilon \to 0^+$ is understood. 
The $\epsilon$-prescription specifies the analytic continuation of the hypergeometric function from 
$\mathcal{Z}<1$ to $\mathcal{Z}>1$ across the logarithmic branch point at $\mathcal{Z}=1$, 
as follows from the Hadamard property of $|0_E\rangle$, 
reviewed in \cite{Epstein:2020wgf,Higuchi:2022nfy}. 

Let now $(\eta,\chi)$ be the coordinates in the metric \eqref{eq:coshchart-metric} on standard $dS_2$, with $(\eta,\chi) \sim (\eta,\chi + 2\pi)$. 
On~$M_p$, we may use the same coordinates but with the identification $(\eta,\chi) \sim (\eta,\chi + 2\pi/p)$. 
It then follows that 
\begin{subequations}
\begin{align}
G_{M_p}(\eta,\chi; \eta',\chi')
& = 
G_{dS_2}\bigl(\mathsf{x}(\eta,\chi), \mathsf{x}(\eta',\chi')\bigr) 
+ 
\Delta G_{M_p}(\eta,\chi; \eta',\chi')
\,,
\\
\Delta G_{M_p}(\eta,\chi; \eta',\chi')
& = 
\sum_{r=1}^{p-1}G_{dS_2}\bigl(\mathsf{x}(\eta,\chi), \mathsf{x}(\eta',\chi' + 2\pi r/p)\bigr) 
\,. 
\end{align}
\label{WightmanFunct_p-quotient}
\end{subequations}

\subsection{Sinh scale factor FRW\label{sec:sinhchart-qft}}

For the sinh scale factor spacetime~\eqref{eq:sinhchart-metric}, 
with $\chi_1$ periodic with period $\lambda_1 \in (0,\infty)$, the formalism of 
Section \ref{CompactifiedCosmol} applies with 
$L = \lambda_1$ and $C(\eta_1) = {(H\sinh \eta_1)}^{-2}$. 

A pair of linearly independent solutions to the mode function differential equation 
\eqref{ScalarFrequency_ode-discrete} are a suitable 
choice of two from 
$\sqrt{\sinh(-\eta_1)} \, P^{\pm\nu}_{-\frac{1}{2}+ i|k_n|}(\cosh\eta_1)$, 
$\sqrt{\sinh(-\eta_1)} \, \bm{Q}^{\pm\nu}_{-\frac{1}{2}+ i|k_n|}(\cosh\eta_1)$
and 
$\sqrt{\sinh(-\eta_1)} \,\, \overline{\bm{Q}^{\pm\nu}_{-\frac{1}{2}+ i|k_n|}(\cosh\eta_1)}$, 
where 
$P^{\pm\nu}_{-\frac{1}{2}+ i|k_n|}$ and 
$\bm{Q}^{\pm\nu}_{-\frac{1}{2}+ i|k_n|}$ are the 
associated Legendre functions at real argument greater than~$1$, 
in the notation of~\cite{dlmf}, and the overline denotes the complex conjugate.

\subsubsection{$n\ne0$ modes}

At $\eta_1 \to -\infty$, 
the $n\ne0$ modes have an adiabatic ``in" vacuum satisfying \eqref{pos:frequency_ScalarMode_nonstatic}. 
The mode functions are  
\begin{align}
\label{AdibaticVacMoFuct:open-spatial-section}
\chi_n^{in}(\eta_1) = 
\frac{\sqrt{\sinh(-\eta_1)}}{\sqrt{\sinh(\pi|k_n|)}}
 \, \overline{\bm{Q}^\nu_{-\frac{1}{2} + i|k_n|}(\cosh\eta_1)} \, ,
\end{align}
as can be verified using 14.3.7 and 14.3.10 in~\cite{dlmf}. 
An alternative expression, 
found via connection formula 13 on p158 of~\cite{MOS1966}, is 
\begin{align}
\chi_n^{in}(\eta_1) 
&= 
\frac{e^{i\varphi_n} e^{-i |k_n| \eta_1}}{\sqrt{2|k_n|}}
\, 
{}_2 F_1 
\! 
\left(
\tfrac{1}{2} + \nu, \tfrac{1}{2} - \nu;1 - i|k_n|; 
\frac{-1}{e^{-2\eta_1} - 1} 
\right)\, ,
\label{AdibaticVacMoFuct:open-spatial-section-alt}
\end{align}
where $\varphi_n$ is a real-valued $n$-dependent phase constant, 
expressible in terms of Euler's gamma-function. 
At large~$|n|$, \eqref{AdibaticVacMoFuct:open-spatial-section-alt}
shows that the mode functions $\chi_n^{in}(\eta_1)$ reduce to the standard plane waves, 
$\exp(-i|k_n|\eta_1)(2|k_n|)^{-1/2}$, 
and the state defined by these 
modes is in this sense an adiabatic vacuum at all times. 

The $n\ne0$ ``in'' vacuum contribution to the 
Wightman function is given by \eqref{eq:Gosc-pre} with 
\eqref{AdibaticVacMoFuct:open-spatial-section-alt} and $L = \lambda_1$.

\subsubsection{Zero momentum mode}

The zero momentum mode is a massive zero mode at $\eta_1\to-\infty$, 
and the early time behaviour does not single out a distinguished mode function. 
We choose a distinguished mode function by matching to the late time behaviour of the Euclidean vacuum modes 
\eqref{eq:Evac-exp-modes-as-nugen} and~\eqref{eq:Evac-exp-modes-as-nu0}
with $n=0$, while continuing to label these modes by the superscript ``in'': 
the outcome for $\nu\ne0$ is 
\begin{align}
\chi_0^{in} (\eta_1)
&= 
\frac{\sqrt{\cos(\pi\nu)} \sqrt{\sinh(-\eta_1)}}{2 \sin(\pi\nu)}
\notag 
\\
& 
\hspace{3ex}
\times 
\left[
e^{i\pi\nu/2} \Gamma\bigl(\tfrac12-\nu\bigr) P^{-\nu}_{-\frac{1}{2}}(\cosh\eta_1)
- 
e^{-i\pi\nu/2} \Gamma\bigl(\tfrac12+\nu\bigr) P^{\nu}_{-\frac{1}{2}}(\cosh\eta_1)
\right] 
\,,
\label{eq:sinh-latetime-zeromode-nugen}
\end{align}
and the corresponding expression for $\nu=0$ is 
\begin{align}
\chi_0^{in}(\eta_1)
= 
\sqrt{\sinh(-\eta_1)}
\left[
\tfrac12 \sqrt\pi \, i 
P_{-\frac{1}{2}}(\cosh\eta_1)  
+ \bm{Q}_{-\frac{1}{2}}(\cosh\eta_1)
\right] \,, 
\label{eq:sinh-latetime-zeromode-nu0}
\end{align}
as can be verified using 
14.2.9 and 14.8.9 in~\cite{dlmf}. 
The zero momentum mode contribution to the 
Wightman function is given by 
\eqref{eq:Gnought-pre} with $L=\lambda_1$.

\subsubsection{Euclidean quotient state\label{sec:sinh-chart-quotient-state}}

We shall not attempt to define an ``out'' vacuum on the sinh scale factor FRW spacetime. 
We shall however consider on the sinh scale factor FRW spacetime a state that can be regarded as a minimal modification of the Euclidean vacuum on~$dS_2$: 
the state whose Wightman function comes from that of the Euclidean vacuum by the periodic image sum in~$\chi_1$. 
We call this state the Euclidean quotient state. 

Proceeding as with the finite quotients in Section~\ref{sec:coshchart-quantisation}, 
the Wightman function $G_{EQ}$ in the Euclidean quotient state reads 
\begin{subequations}
\label{WightmanFunct_EQ}
\begin{align}
& G_{EQ}(\eta_1,\chi_1; \eta_1',\chi_1')
= 
G_{dS_2}\bigl(\mathsf{x}(\eta_1,\chi_1), \mathsf{x}(\eta_1',\chi_1')\bigr) 
+ 
\Delta G_{EQ}\bigl(\mathsf{x}(\eta_1,\chi_1), \mathsf{x}(\eta_1',\chi_1')\bigr) 
\,,
\\
& \Delta G_{EQ}\bigl(\mathsf{x}(\eta_1,\chi_1), \mathsf{x}(\eta_1',\chi_1')\bigr) 
= 
\sum_{r \in \BbbZ\setminus\{0\}}G_{dS_2}
\bigl(\mathsf{x}(\eta_1,\chi_1), \mathsf{x}(\eta_1',\chi_1' + r\lambda_1)\bigr) 
\,, 
\label{WightmanFunct_EQ-imageterms}
\end{align}
\end{subequations}
where $G_{dS_2}$ is given by \eqref{WightmanFunct_Standard_dS} and~\eqref{eq:Z-standard-dS}. 
It follows from the embedding equations \eqref{eq:sinhchart-embedding}, 
the definition \eqref{RealQuadraticForm} and the asymptotics of the hypergeometric function in \eqref{WightmanFunct_Standard_dS} that the sum in \eqref{WightmanFunct_EQ-imageterms} converges exponentially, making $G_{EQ}$ well defined. The coincidence limit singularity of $G_{EQ}$ is Hadamard because this singularity comes from~$G_{dS_2}$. 

The Euclidean quotient state is however not a pure state in the Fock space description of Section~\ref{subsec:untwisted-theory}.  
This is because $dS_2$ contains two opposing future sinh chart patches before the periodic identification: 
the induced state on the union of the two patches 
contains correlations between the two patches, 
much like the more familiar correlations between the two opposing static coordinate patches~\cite{Lapedes:1977ip}, 
as can be analysed by extending the future sinh chart techniques of \cite{Louko:1998qf} from their $dS_4/\BbbZ_2$ context to the present context on~$dS_2$. 

We shall employ the image sum expression \eqref{WightmanFunct_EQ}
in Sections \ref{sec:comoving-detectors} and \ref{sec:stress-energy} below.

\subsection{Exp scale factor FRW\label{sec:exp-chart-QFT}}

For the exp scale factor spacetime~\eqref{eq:expchart-metric}, 
with $\chi_0$ periodic with period $\lambda_0 \in (0,\infty)$, the formalism of 
Section \ref{CompactifiedCosmol} applies with 
$L = \lambda_0$ and $C(\eta_0) = {(H\eta_0)}^{-2}$. 

For $k_n\ne0$, a linearly independent pair of 
solutions to the mode function differential equation 
\eqref{ScalarFrequency_ode-discrete} are 
$\sqrt{-\eta_0} \, H^{(1)}_\nu(-|k_n|\eta_0)$
and $\sqrt{-\eta_0} \, H^{(2)}_\nu(-|k_n|\eta_0)$, 
where $H^{(1,2)}_\nu$ are the Hankel functions~\cite{dlmf}. 
For $k_n =0$, a linearly independent pair is 
${(-\eta_0)}^{\frac{1}{2}\pm \nu}$ when $\nu\ne0$, 
and 
${(-\eta_0)}^{\frac{1}{2}}$ and ${(-\eta_0)}^{\frac{1}{2}}\ln(-\eta_0)$ when $\nu=0$.

\subsubsection{$n\ne0$ modes}

At $\eta_0 \to -\infty$, 
the $n\ne0$ modes admit an adiabatic ``in" form 
satisfying~\eqref{pos:frequency_ScalarMode_nonstatic}, 
with the mode functions 
\begin{align}
\chi_n^{in}(\eta_0) = 
\tfrac12 \sqrt{\pi} \, e^{i\pi\nu/2}
\, 
\sqrt{-\eta_0} \, H^{(1)}_\nu(-|k_n|\eta_0) \, . 
\label{eq:Exp-osc-inmodes}
\end{align}
These mode functions reduce to the standard plane waves, 
$\exp(-i|k_n|\eta_0)(2|k_n|)^{-1/2}$, 
also in the limit of large $|n|$ with fixed~$\eta_0$, and the 
``in'' vacuum defined by these 
modes is in this sense an adiabatic vacuum at all times~\cite{Birrell:1982ix}. 
Were $\chi_0$ not periodic, the ``in'' vacuum defined by the corresponding modes, 
with $k_n$ replaced by a continuous momentum, 
would coincide with the restriction of the 
Euclidean vacuum to the exp chart on $dS_2$~\cite{Bunch:1978yq}. 

The $n\ne0$ ``in'' vacuum contribution to the 
Wightman function is given by \eqref{eq:Gosc-pre} with 
\eqref{eq:Exp-osc-inmodes} and $L = \lambda_0$. 
Note that this contribution is invariant under redefinitions of $\lambda_0$ 
by the coordinate rescalings~\eqref{eq:expchart-rescaling}. 

While we shall not attempt to define distinguished late time modes, 
we note that at late times, $\eta_0 \to 0^-$, 
the leading two terms in the asymptotic behaviour of $\chi_n^{in}$ 
\eqref{eq:Exp-osc-inmodes} for $\nu\ne0$ are 
\begin{align}
\chi_n^{in} (\eta_0)
&\sim 
\frac{- i \sqrt\pi \, {(-\eta_0)^{1/2}}}{2 \sin(\pi\nu)}
\left[
\frac{ e^{i\pi\nu/2} \, 2^\nu {|k_n|}^{-\nu}}{\Gamma(1-\nu)} 
{(-\eta_0)}^{-\nu}
- 
\frac{ e^{-i\pi\nu/2} \, 2^{-\nu} {|k_n|}^{\nu}}{\Gamma(1+\nu)} 
{(-\eta_0)}^{\nu}
\right] 
\,,
\label{eq:Evac-exp-modes-as-nugen}
\end{align}
and the expression for $\nu=0$ is 
\begin{align}
\chi_n^{in} (\eta_0)
\sim 
\frac{-i {(-\eta_0)}^{1/2}}{\sqrt\pi}
\left[
\tfrac12 \pi i  + \ln2 - \gamma - \ln(|k_n|) - \ln(-\eta_0)
\right] \,. 
\label{eq:Evac-exp-modes-as-nu0}
\end{align}
The $\eta_0$-dependence in 
\eqref{eq:Evac-exp-modes-as-nugen} and \eqref{eq:Evac-exp-modes-as-nu0} 
is similar to the $(\pi-\eta)$-dependence in 
\eqref{eq:Evac-modes-as-nugen} and~\eqref{eq:Evac-modes-as-nu0}, 
but the $|k_n|$-dependence in the coefficients is not directly comparable, 
due to the differing foliations in the $dS_2$ embedding spacetime. 


\subsubsection{Zero momentum mode}

The zero momentum mode is a massive zero mode at $\eta_0\to-\infty$, 
and the early time behaviour does not single out a distinguished mode function. 
We choose a distinguished mode function by the late time behaviour, 
matching the $\eta_0 \to 0^-$ asymptotics with $\lambda_0 = 2\pi$ to the $\eta\to\pi^-$ 
asymptotics of the Euclidean vacuum $n=0$ mode function~\eqref{ModeFunc-ClosedSPS}, 
using \eqref{eq:Evac-modes-as-nugen} and~\eqref{eq:Evac-modes-as-nu0}, 
and including a $\lambda_0$-dependent scaling of $\eta_0$ to make the contribution to the Wightman function invariant under redefinitions of $\lambda_0$ by the coordinate rescalings~\eqref{eq:expchart-rescaling}. 
For $\nu\ne0$, we obtain 
\begin{align}
& \chi_0^{in}(\eta_0) 
= 
\frac{\sqrt{\cos(\nu\pi)} \, {(-\eta_0)}^{1/2}}{2\sin(\nu\pi)} 
\notag 
\\[1ex]
&\hspace{3ex}
\times \left[ 
\frac{ e^{i\pi\nu/2} \, 2^{\nu} \Gamma\left( \frac{1}{2} - \nu \right) {(2\pi/\lambda_0)}^{-\nu} }{\Gamma(1 -\nu)}
\, 
{(-\eta_0)}^{-\nu}
- 
\frac{e^{-i\pi\nu/2} \, 2^{-\nu} \Gamma\left( \frac{1}{2} + \nu \right) {(2\pi/\lambda_0)}^{\nu}}{\Gamma(1 +\nu)}
\,
{(-\eta_0)}^{\nu}
\right] 
\,, 
\label{eq:expchart-latetimemode-nugen}
\end{align}
and the expression for $\nu=0$ is 
\begin{align}
\chi_0^{in}(\eta_0) 
=  
\frac{{(-\eta_0)}^{1/2}}{\sqrt\pi}
\left[
\tfrac12 \pi i + \ln(4\lambda_0/\pi) - \ln(-\eta_0) 
\right] \,. 
\label{eq:expchart-latetimemode-nu0}
\end{align}
The zero momentum mode contribution to the 
Wightman function is given by 
\eqref{eq:Gnought-pre} with $L=\lambda_0$, 
and is invariant under redefinitions of $\lambda_0$ by~\eqref{eq:expchart-rescaling}.

\subsubsection{Euclidean quotient state?\label{sec:expchart-quotientstate}}

On the exp scale factor FRW spacetime, an attempt to define the 
Wightman function of an Euclidean quotient state by adapting the image sum in \eqref{WightmanFunct_EQ} 
to the periodic identification in $\chi_0$ produces an image sum that is not convergent in absolute value, due to the weak falloff of the hypergeometric function. 
We shall see in Section \ref{sec:expchart-stressenergy} 
that a similar weak convergence appears also in the image sum expression for the stress-energy tensor. 
We shall return to this phenomenon in the concluding remarks in Section~\ref{sec:conclusions}.

\section{Comoving detector\label{sec:comoving-detectors}}

In this section we address the response of a comoving UDW detector in the cosh, sinh and exp scale factor FRW spacetimes.

\subsection{Detector and its response function}

We consider a spatially pointlike two-level system known as the Unruh-DeWitt dectector \cite{Unruh:1976db,DeWitt:1980hx}, on a trajectory $\mathsf{x}(\tau)$, parametrised by the proper time~$\tau$, and coupled linearly to the scalar field at the detector's position, $\phi\bigl(\mathsf{x}(\tau)\bigr)$. 

As reviewed in~\cite{Toussaint:2021czo}, 
the detector's transition probability in first-order perturbation theory is a multiple of the response function~$\mathcal{F}(\omega)$, given by 
\begin{align}
\mathcal{F}(\omega) := 
\int
d\tau \, d\tau' \, 
\chi(\tau) \chi(\tau') 
\, e^{-i\omega(\tau-\tau')} \, G(\tau,\tau') \, ,
\label{RespFunct:Scalarfiel_CB}
\end{align}
where $\omega$ is the detector's energy gap, $\chi$ is a switching function that specifies how the interaction is turned on and off, and $G(\tau,\tau')$ is the pull-back of the field's Wightman function to the detector's wordline, 
\begin{align}
G (\tau, \tau')
:= 
\langle \Psi | 
\hat{\phi}\bigl( \mathsf{x}(\tau) \bigr)
\hat{\phi}\bigl(\mathsf{x}(\tau') \bigr)
|\Psi \rangle
\, , 
\end{align}
with $|\Psi\rangle$ being the state in which the field was prepared before the interaction starts. Positive values of $\omega$ give the probability of excitation and negative values of $\omega$ give the probability of de-excitation. 

When $|\Psi \rangle$ is a Hadamard state, $G (\tau, \tau')$ is a well-defined distribution~\cite{Hormander:1983,Fewster:1999gj}, 
and $\mathcal{F}(\omega)$ is well defined whenever $\chi$ is smooth and has compact support. 
In our present case of $1+1$ spacetime dimensions, 
the coincidence limit singularity of $G (\tau, \tau')$ is only logarithmic, and hence integrable, 
and $\mathcal{F}(\omega)$ remains well defined even for less regular~$\chi$. 
We use this freedom to adopt a $\chi$ that has minimal structure of its own, 
having a constant value between a sharp switch-on moment and a sharp switch-off moment; 
we choose 
\begin{align}
\chi(\tau) = \Theta (\tau - \tau_0) \Theta(\tau_1 - \tau) 
\, ,
\end{align}
where $\tau_0$  and $\tau_1$ denote the switch-on and switch-off moments, respectively, 
and we assume $\tau_0 < \tau_1$. 
The response function \eqref{RespFunct:Scalarfiel_CB} then becomes
\begin{align}
\label{ResponsFunct_2D}
\mathcal{F}(\omega, \tau_1, \tau_0)
:= 
\int _{\tau_0}^{\tau_1}
d\tau \, \int _{\tau_0}^{\tau_1} d\tau' \, 
\, e^{-i\omega(\tau-\tau')} \, G(\tau,\tau')
\, .
\end{align}

We specialise to a comoving detector, at a constant value of $x$ in the FRW coordinates $(t,x)$ of~\eqref{ExpComosMetric}. 
As all the quantum states that we consider are invariant under shifts in the compactified spatial coordinate~$x$, 
we may without loss of generality take the trajectory to be $\bigl(t(\tau),x(\tau)\bigr) = (\tau,0)$.

\subsection{Cosh scale factor FRW}

In the cosh scale factor FRW spacetime, 
for general~$\lambda$, 
a mode sum expression for the response function 
is obtained from \eqref{eq:G-totalsum}, \eqref{ModeFunc-ClosedSPS-alt} and~\eqref{ResponsFunct_2D}, with the outcome 
\begin{align}
\label{ResponsFunct-Coshchart-General_Form}
& \mathcal{F}(\omega, \tau_1, \tau_0) 
= \frac{1}{2\lambda}
\sum_{n=-\infty}^\infty 
\frac{\Gamma\bigl(\tfrac12 + \nu + |k_n|\bigr)
\Gamma\bigl(\tfrac12 - \nu + |k_n|\bigr)}{{\bigl(\Gamma(1 + |k_n|)\bigr)}^2}
\notag
\\
&\hspace{5ex}
\times 
\left| \int_{\tau_0}^{\tau_1}
d\tau \, 
e^{-i\omega\tau - 2i|k_n|\arctan(e^{H\tau})} 
{}_2 F_1 
\Bigl(
\tfrac{1}{2} + \nu, \tfrac{1}{2} - \nu;1 + |k_n|; 
\tfrac12 \bigl(1 - i \sinh(H\tau)\bigr) 
\Bigr)
\right|^2\, ,
\end{align}
where we recall that $k_n = 2\pi n/\lambda$. 
We have used \eqref{ModeFunc-ClosedSPS-alt} in preference to \eqref{ModeFunc-ClosedSPS}
as this is likely to be more suitable for numerical evaluation of the large $|n|$ terms. 
The contributions from $n>0$ and $n<0$ are equal and can be combined to a single sum over $n$ from $1$ to~$\infty$. 

On standard $dS_2$, in which $\lambda = 2\pi$, 
and on its $M_p$ quotients, in which $\lambda = 2\pi/p$, $p = 2,3,\ldots$, 
an alternative is to use the Wightman function as given in 
\eqref{WightmanFunct_Standard_dS} and~\eqref{WightmanFunct_p-quotient}. 
Setting $M_1 = dS_2$, a formula covering $M_p$ for all $p = 1,2,\ldots$ is 
\begin{align}
\label{ResponsFunct-Coshchart}
\mathcal{F}_p(\omega, \tau_1, \tau_0)
=\tfrac14 
\sec(\pi \nu) 
\sum_{r=0}^{p-1}
\int _{\tau_0}^{\tau_1}
d\tau \, \int_{\tau_0}^{\tau_1} d\tau' \, 
\, e^{-i\omega(\tau-\tau')} \, 
{}_2F_1 \! \left(\tfrac{1}{2} - \nu, \tfrac{1}{2} + \nu;1; g_{p,r}(\tau,\tau')\right)
\, ,
\end{align}
where
\begin{align}
g_{p,r}(\tau,\tau') :=
\cosh^2\bigl(\tfrac12 H(\tau-\tau') - i \epsilon\bigr)
- \cosh(H \tau - i \epsilon)\cosh(H \tau' + i \epsilon)\sin^2(r \pi/p)
\label{eq:gpr-definition}
\end{align}
and the limit $\epsilon\to0^+$ is understood; 
as in \eqref{WightmanFunct_Standard_dS}
and~\eqref{eq:Z-standard-dS}, 
the $i\epsilon$ prescription specifies the analytic continuation of 
${}_2F_1$ from $g_{p,r}<1$ to $g_{p,r}>1$ across the logarithmic singularity at $g_{p,r}=1$. 
Note that for the pure $dS_2$ term, $r=0$, only the first term in \eqref{eq:gpr-definition} is present.

\subsection{Sinh scale factor FRW}

\subsubsection{``In'' vacuum}

In the sinh scale factor FRW spacetime, the contribution to the response function from 
$n\ne0$ modes in the ``in'' vacuum 
is obtained from \eqref{eq:Gosc-pre}, 
\eqref{AdibaticVacMoFuct:open-spatial-section-alt} and~\eqref{ResponsFunct_2D}, 
with the outcome 
\begin{align}
& \mathcal{F}^{in}_{osc}(\omega, \tau_1, \tau_0) 
= \frac{1}{2\lambda_1}
\sum_{n=-\infty}^\infty 
\frac{1}{|k_n|}
\notag
\\
&\hspace{5ex}
\times 
\left| 
\int_{\tau_0}^{\tau_1}
d\tau \, 
e^{-i\omega\tau - i|k_n|\ln \tanh(e^{H\tau/2})} 
\, 
{}_2 F_1 
\Bigl(
\tfrac{1}{2} + \nu, \tfrac{1}{2} - \nu;1 -i|k_n|; 
- \sinh^2(H\tau/2) 
\Bigr)
\right|^2\, ,
\end{align}
where we recall that now $k_n = 2\pi n/\lambda_1$. 
The contributions from $n>0$ and $n<0$ are again 
equal and can be combined to a single sum over $n$ from $1$ to~$\infty$. 

The contribution from the $n=0$ mode in the ``in'' vacuum is obtained from 
\eqref{eq:Gnought-pre}, 
\eqref{eq:sinh-latetime-zeromode-nugen}, 
\eqref{eq:sinh-latetime-zeromode-nu0}
and~\eqref{ResponsFunct_2D}. 
For $\nu\ne0$, the outcome is 
\begin{align}
\mathcal{F}^{in}_{0}(\omega, \tau_1, \tau_0) = 
\frac{1}{\lambda_1}\left( |N_+|^2 + |N_-|^2 + 2 \Realpart(N_+ N_-^*) \right) \, ,
\end{align}
where 
\begin{align}
N_{\pm} = 
\pm \frac{ e^{\pm i\pi\nu/2} \sqrt{\cos(\pi\nu)} \, \Gamma\bigl(\tfrac12\mp\nu\bigr)}{2 \sin(\pi\nu)}
\int_{\tau_0}^{\tau_1}   
\frac{ d\tau \, e^{- i\omega\tau } }{\sqrt{\sinh(H\tau) }} P^{\mp \nu}_{-\frac12} \bigl(\coth(H\tau)\bigr)\, ,
\end{align}
and for $\nu = 0$ 
the corresponding expression is 
\begin{align}
\mathcal{F}^{in}_{0}(\omega, \tau_1, \tau_0) = \frac{1}{\lambda_1}
\left|
\int_{\tau_0}^{\tau_1}   \frac{ d\tau \, e^{- i\omega\tau } }{\sqrt{\sinh(H\tau) }} 
\left(
\tfrac12 \sqrt\pi \, i P_{-\frac12} \bigl(\coth(H\tau)\bigr) 
+ \bm{Q}_{-\frac12} \bigl(\coth(H\tau)\bigr)
\right)
\right|^2
\,. 
\end{align}

\subsubsection{Euclidean quotient state}

In the Euclidean quotient state, proceeding as with \eqref{ResponsFunct-Coshchart} gives the response function 
\begin{align}
\label{ResponsFunct-sinhchart-EQ}
\mathcal{F}_{EQ}(\omega, \tau_1, \tau_0)
=\tfrac14 
\sec(\pi \nu) 
\sum_{r\in\BbbZ}
\int _{\tau_0}^{\tau_1}
d\tau \, \int_{\tau_0}^{\tau_1} d\tau' \, 
\, e^{-i\omega(\tau-\tau')} \, 
{}_2F_1 \! \left(\tfrac{1}{2} - \nu, \tfrac{1}{2} + \nu;1; g_r(\tau,\tau')\right)
\, ,
\end{align}
where
\begin{align}
g_r(\tau,\tau') :=
\cosh^2\bigl(\tfrac12 H(\tau-\tau') - i \epsilon\bigr)
- \sinh(H \tau - i \epsilon)\sinh(H \tau' + i \epsilon)\sinh^2(r\lambda_1/2)
\label{eq:gr-definition}
\end{align}
and the limit $\epsilon\to0^+$ is again understood. The $r=0$ term is the pure $dS_2$ contribution.

\subsection{Exp scale factor FRW\label{sec:exp-chart-detector}}

In the exp scale factor FRW spacetime, 
the contribution to the response function from the $n\ne0$ modes in the ``in'' vacuum is obtained from 
from \eqref{eq:Gosc-pre}, 
\eqref{eq:Exp-osc-inmodes}
and~\eqref{ResponsFunct_2D}, 
with the outcome 
\begin{align}
\mathcal{F}^{in}_{osc}(\omega, \tau_1, \tau_0) 
&= 
\frac{\pi 
e^{i\pi(\nu-\nu^*)/2}}{2\lambda_0}\sum_{n=1}^\infty \left| \int_{\tau_0}^{\tau_1}   d\tau \, e^{-\left(\frac{H}{2} + i\omega \right)\tau} 
H^{(1)}_\nu\left(\frac{2\pi n}{\lambda_0} e^{-H\tau} \right) \right|^2
\notag
\\[1ex]
&= 
\frac{\pi 
e^{i\pi(\nu-\nu^*)/2}}{2\lambda_0 e^{H\tau_0}}\sum_{n=1}^\infty \left| \int_0^{\tau_1-\tau_0} ds \, e^{-\left(\frac{H}{2} + i\omega \right)s} 
H^{(1)}_\nu\left(\frac{2\pi n}{\lambda_0 e^{H\tau_0}} e^{-Hs} \right) \right|^2
\,,
\label{eq:expchart-oscmodes-response-final}
\end{align}
where the last expression is obtained by the substitution $\tau = \tau_0+s$. 
The last expression in \eqref{eq:expchart-oscmodes-response-final} shows that 
$\mathcal{F}^{in}_{osc}$ depends on $\tau_0$, $\tau_1$ and $\lambda_0$, none of which has a geometrically invariant meaning, 
only through the combinations $\lambda_0 e^{H\tau_0}$ and $\tau_1-\tau_0$, which do: 
$\lambda_0 e^{H\tau_0}/H$ is the spatial circumference at the moment of cosmological time when the detector is turned on, and $\tau_1-\tau_0$ is the proper time that the detector operates. 

The contribution from the $n=0$ mode in the ``in'' vacuum is obtained from 
\eqref{eq:Gnought-pre}, 
\eqref{eq:expchart-latetimemode-nugen}, 
\eqref{eq:expchart-latetimemode-nu0}
and~\eqref{ResponsFunct_2D}. 
For $\nu\ne0$, the outcome is 
\begin{align}
\mathcal{F}^{in}_{0}(\omega, \tau_1, \tau_0) = \frac{1}{\lambda_0}\left( |M_+|^2 + |M_-|^2 + 2 \Realpart(M_+ M_-^*) \right) \, ,
\label{eq:expchart-zeroF-nugen-sum}
\end{align}
where 
\begin{align}
M_{\pm} &: = \pm 
\frac{ e^{\pm i\pi\nu/2} \, 2^{\pm\nu -1} \sqrt{\cos(\nu\pi)} \, \Gamma\left( \frac{1}{2} \mp \nu \right) {(2\pi/\lambda_0)}^{\mp\nu}}{\sin(\nu\pi)\Gamma(1 \mp\nu)}
\int_{\tau_0}^{\tau_1}
d\tau \, e^{-\left(\frac{1}{2} \mp\nu\right)H\tau -i\omega \tau}
\notag
\\
& \;= \pm 
\frac{ e^{\pm i\pi\nu/2} \, 2^{\pm\nu -1} \sqrt{\cos(\nu\pi)} \, \Gamma\left( \frac{1}{2} \mp \nu \right) 
{\bigl(2\pi /(\lambda_0 e^{H\tau_0})\bigr)}^{\mp\nu} e^{-i\omega \tau_0}}{\sin(\nu\pi)\Gamma(1 \mp\nu) e^{H\tau_0/2}}
\notag
\\
&\hspace{5ex}
\times 
\int_0^{\tau_1-\tau_0}
ds \, e^{-\left(\frac{1}{2} \mp\nu\right)Hs -i\omega s}
\,,
\label{eq:expchart-zeroF-nugen-M}
\end{align}
and for $\nu=0$ the corresponding expression is 
\begin{align}
\mathcal{F}^{in}_{0}(\omega, \tau_1, \tau_0) 
& =
\frac{1}{\pi\lambda_0}
\left|
\int_{\tau_0}^{\tau_1} d\tau \, 
e^{-\left(\frac{H}{2} + i\omega\right)\tau}
\Bigl(
\tfrac12 \pi i + \ln(4\lambda_0/\pi) + H\tau 
\Bigr)
\right|^2
\notag
\\
& =
\frac{1}{\pi\lambda_0 e^{H\tau_0}}
\left|
\int_0^{\tau_1-\tau_0}   ds \, 
e^{-\left(\frac{H}{2} + i\omega\right)s}
\Bigl(\tfrac12 \pi i + \ln\bigl(4\lambda_0 e^{H\tau_0} /\pi\bigr) + Hs \Bigr)
\right|^2
\,. 
\label{eq:expchart-zeroF-nuzero}
\end{align}
From the last expressions in \eqref{eq:expchart-zeroF-nugen-M}
and 
\eqref{eq:expchart-zeroF-nuzero}
it is clear that $\mathcal{F}^{in}_0$ depends on $\tau_0$, $\tau_1$ and $\lambda_0$
only through the geometrically invariant combinations $\lambda_0 e^{H\tau_0}$ and $\tau_1-\tau_0$.

\section{Stress-energy tensor\label{sec:stress-energy}}

In this section we evaluate the renormalised stress-energy tensor 
in the states whose Wightman function comes from that in $|0_E\rangle$ 
on $dS_2$ by an image sum construction. 

\subsection{Point-splitting and image sum}

Recall that the classical stress-energy tensor for a massive scalar field in 
$1+1$ dimensions reads \cite{Birrell:1982ix}
\begin{align}
T_{\mu\nu} 
&= 
(1-2\xi)\phi_{,\mu}\phi_{,\nu} + \left(2\xi - \tfrac{1}{2}\right) \! g_{\mu\nu}g^{\rho\sigma}\phi_{,\rho}\phi_{,\sigma}  - 2\xi\phi\phi_{;\mu\nu} +
\xi g_{\mu\nu}\phi\Box\phi 
\notag 
\\[1ex]
& \quad
+ \Bigl[\left(\tfrac{1}{2} - \xi\right) \! m^2 - \xi^2 R \Bigr] g_{\mu\nu}\phi^2\, .
\label{eq:Tmunu-reallygeneral}
\end{align}
The point-splitting definition of the renormalised expectation value of $T_{\mu\nu}$
in a quantum state $|\Psi\rangle$ is 
\begin{align}
\langle \Psi | T_{\mu\nu}(\mathsf{x})|\Psi\rangle 
&:= \lim_{\mathsf{x}\rightarrow \mathsf{x}'} \mathcal{D}_{\mu\nu}(\mathsf{x},\mathsf{x}' ) G^{(1)}_{ren}(\mathsf{x},\mathsf{x}' )
\,, 
\label{eq:T-ren-general}
\end{align}
where 
$\mathcal{D}_{\mu\nu}\left(\mathsf{x},\mathsf{x}'\right)$ 
is a two-point differential operator obtained by promoting the differential operators acting at a single spacetime point in \eqref{eq:Tmunu-reallygeneral} into suitably point-split versions, 
and $G^{(1)}_{ren}(\mathsf{x},\mathsf{x}')$ is obtained from the Hadamard function, 
$G^{(1)}(\mathsf{x},\mathsf{x}'):= G(\mathsf{x},\mathsf{x}') + G(\mathsf{x}',\mathsf{x})$, 
where $G(\mathsf{x},\mathsf{x}') = \langle \Psi | \phi(\mathsf{x}) \phi(\mathsf{x}') | \Psi\rangle$, 
by subtracting purely geometric state-independent counterterms~\cite{Birrell:1982ix,decanini-folacci}. 

We shall consider states induced 
by a quotient construction from $|0_E\rangle$ on~$dS_2$. In these states, 
renormalisation is required only in the $|0_E\rangle$ contribution on~$dS_2$, 
with the well-known outcome \cite{Bunch:1978yq}
\begin{align}
\label{Contribution-dS}
\langle 0_E| T_{\mu\nu} |0_E\rangle 
&= 
-\frac{1}{8\pi} \Biggl\lbrace m^2\ln \! \left(\frac{H^2}{m^2}\right) 
+ m^2 \! \left[ \psi \! \left(\tfrac{1}{2} +\nu \right) 
+ \psi \! \left(\tfrac{1}{2} -\nu \right) \right] 
+ \left(\tfrac{1}{3} - 2\xi \right) \! H^2 \Biggr\rbrace 
g_{\mu\nu}
\, , 
\end{align}
where $\psi$ is the digamma function~\cite{dlmf}. 
The additional contributions from the image terms require no renormalisation, 
and can be evaluated with minimal technology, as follows. 

First, observe that in a locally de~Sitter spacetime, 
where $R = 2H^2$, 
and using the field equation, 
$(\Box + m^2 + 2\xi H^2 )\phi =0$, 
the classical expression
\eqref{eq:Tmunu-reallygeneral} for 
$T_{\mu\nu}$ reduces to 
\begin{align}\label{Classical-SEM-tensor}
T_{\mu\nu} &= (1-2\xi)\phi_{,\mu}\phi_{,\nu} 
+ \left(2\xi - \tfrac{1}{2}\right) \! g_{\mu\nu}g^{\rho\sigma}\phi_{,\rho}\phi_{,\sigma}  - 2\xi\phi\phi_{;\mu\nu} 
\notag 
\\[1ex]
&\quad
+ \Bigl[ \left(\tfrac{1}{2}-2\xi\right) \! m^2 - 4\xi^2H^2\Bigr] g_{\mu\nu}\phi^2 \, .
\end{align} 
For the two-point differential operator in~\eqref{eq:T-ren-general}, 
for the image terms we may therefore replace ${\mathcal{D}}_{\mu\nu}$ by 
\begin{align}
{\tilde{\mathcal{D}}}_{\mu\nu}\left(\mathsf{x},\mathsf{x}'\right)
& = 
\tfrac{1}{4}(1-2\xi)(\nabla_\mu\nabla_{\nu'}  +\nabla_{\mu'} \nabla_\nu) 
+ \tfrac{1}{4} \! \left(2\xi-\tfrac{1}{2} \right) \! g_{\mu\nu}(g^{\rho\sigma'}\nabla_\rho\nabla_{\sigma'}  + g^{\rho'\sigma}\nabla_{\rho'} \nabla_\sigma)
\notag 
\\[1ex]
&\quad 
-\tfrac{1}{2}\xi (\nabla_\mu\nabla_{\nu}  +\nabla_{\mu'} \nabla_{\nu'})
+ \Bigl[ \left(\tfrac{1}{4}-\xi\right) \! m^2 - 2\xi^2H^2\Bigr] g_{\mu\nu} 
\, , 
\label{eq:Dforus-formula}
\end{align}
and, again for the image terms, we need not specify at which point the metric tensors in \eqref{eq:Dforus-formula} are evaluated. 

Second, recall from Sections \ref{sec:coshchart-quantisation}
and \ref{sec:sinhchart-qft} that we now consider the cosh scale factor FRW spacetimes with 
$\lambda = 2\pi/p$ with $p=2,3,...$, 
denoted by~$M_p$, and on them the induced vacuum~$|0_p\rangle$, 
and the sinh scale factor FRW spacetimes with arbitrary~$\lambda_1$, 
and on them the Euclidean quotient state. 
In these states, we have 
\begin{subequations}
\label{eq:G-dS-plus+images}
\begin{align}
G(\mathsf{x},\mathsf{x}') &= G_{dS_2}(\mathsf{x},\mathsf{x}') + \Delta G(\mathsf{x},\mathsf{x}')
\,,
\\
\Delta G(\mathsf{x},\mathsf{x}')
& = 
\tfrac{1}{4}\sec(\pi \nu) \sum_{r \in I} 
F \Bigl(\tfrac{1}{2} \bigl( 1+\mathcal{Z}_r(\mathsf{x}, \mathsf{x}') \bigr)
\Bigr)
\, ,
\label{eq:G-dS-images}
\end{align}
\end{subequations}
where
\begin{align}
F(q) := {}_2 F_1 
\bigl(
\tfrac{1}{2} - \nu, \tfrac{1}{2} + \nu;1; 
q \bigr)
\, ,
\end{align} 
and $\mathcal{Z}_r$ and the index set $I$ depend on the quotient: 
in the $M_p$ spacetime, with $p=2,3,\ldots$, 
working in the chart $(t,\chi)$~\eqref{ClosedSpatialSection-dS}, we have 
\begin{align}
\mathcal{Z}_r \bigl( \mathsf{x}(t,\chi), \mathsf{x}(t',\chi') \bigr)
& =
\cosh(Ht)\cosh(Ht') \cos(\chi-\chi' + 2\pi r/p) - \sinh(Ht)\sinh(Ht')
\,,
\notag
\\
& \hspace{3.3ex}
r \in I = \{ 1,\ldots, p-1\}
\,,
\end{align}
while in the sinh scale factor FRW spacetime, working in the chart 
$(t_1,\chi_1)$~\eqref{OpenSpatialSection-DS}, 
we have 
\begin{align}
\mathcal{Z}_r \bigl( \mathsf{x}(t_1,\chi_1), \mathsf{x}(t_1',\chi_1') \bigr)
& = 
\cosh(Ht)\cosh(Ht_1')  - \sinh(Ht_1)\sinh(Ht_1') \cosh(\chi_1-\chi_1' + r \lambda_1)
\,,
\notag
\\
& \hspace{3.3ex}
r \in I = \BbbZ \setminus \{0\}
\,.
\end{align}
We have dropped the $i\epsilon$ from $\mathcal{Z}_r$ 
because for these image terms $\mathcal{Z}_r(\mathsf{x}, \mathsf{x}') < 1$ 
whenever $\mathsf{x}$ and $\mathsf{x}'$ are sufficiently close, 
which is the case on taking the coincidence limit in~\eqref{eq:T-ren-general}. 

For each of these quotients, we hence have 
\begin{align}
\langle T_{\mu\nu} \rangle 
= 
\langle 0_E| T_{\mu\nu} |0_E\rangle 
+ 
\Delta \langle T_{\mu\nu} \rangle 
\,,
\end{align}
where
\begin{align}
\Delta \langle T_{\mu\nu} \rangle 
= 
\lim_{\mathsf{x}\rightarrow \mathsf{x}'} {\tilde{\mathcal{D}}}_{\mu\nu}(\mathsf{x},\mathsf{x}' ) 
\bigl( 
\Delta G(\mathsf{x},\mathsf{x}') + \Delta G(\mathsf{x}',\mathsf{x})
\bigr) 
\,,
\end{align}
where $\Delta G$ is as given in~\eqref{eq:G-dS-images}. 

We shall now give the expressions for $\Delta \langle T_{\mu\nu} \rangle$. 

\subsection{$M_p$ spacetime}

In the $M_p$ spacetime, 
working in the chart $(t,\chi)$~\eqref{ClosedSpatialSection-dS}, we find that 
$\Delta \langle T_{\mu\nu} \rangle$ is diagonal, with 
\begin{subequations}
\label{Cosh-Chart_VEV}
\begin{align}
\Delta\langle T_{t}{}^t \rangle 
&= 
\tfrac14 H^2\sec(\pi \nu) 
\sum_{r = 1}^{p-1}
\biggl\{
- \xi F(q_r) 
+ \Bigl[ 
\left( 2\xi -\tfrac{1}{2} \right) \! q_r 
-\xi \bigl(1 + \cos(2\pi r/p) \bigr)  
\Bigr] 
F'(q_r)  
\biggr\} 
\,,
\\
- \Delta \langle T_{\chi}{}^\chi \rangle 
&= 
\tfrac14 H^2\sec(\pi \nu) 
\sum_{r = 1}^{p-1}
\biggl\{
\left[(4\xi-1) \! \left(\frac{m^2}{H^2} + 2\xi \right) + \xi \right] \! F(q_r)
\notag
\\
& \hspace{22ex}
+ \Bigl[ 
\left( 4\xi -\tfrac{1}{2} \right) \! q_r 
-\xi \bigl(1 - \cos(2\pi r/p) \bigr)  
\Bigr] 
F'(q_r)
\notag
\\
& \hspace{22ex}
- 2\xi (1 - q_r) \bigl(1 + \cos(2\pi r/p) \bigr) 
F''(q_r)
\biggr\} 
\,,
\end{align}
\end{subequations}
where 
\begin{align}
q_r &:= 1 - \tfrac12 \cosh^2(Ht)\bigl(1 - \cos(2\pi r/p) \bigr) 
\, . 
\end{align}

In the special case $p=2$, the image sum has only one term, $r=1$,
and the expressions in \eqref{Cosh-Chart_VEV} are similar to those for a similar $\BbbZ_2$ quotient of $dS_4$ considered in~\cite{Louko:1998qf}.

\subsection{Sinh scale factor FRW spacetime}

In the sinh scale factor FRW spacetime, 
working in the chart $(t_1,\chi_1)$~\eqref{OpenSpatialSection-DS}, 
we find that 
$\Delta \langle T_{\mu\nu} \rangle$ is diagonal, with 
\begin{subequations}
\label{Sinh-Chart_VEV}
\begin{align}
\Delta\langle T_{t}{}^t \rangle 
&= 
\tfrac12 H^2\sec(\pi \nu) 
\sum_{r = 1}^\infty
\biggl\{
- \xi F({\tilde q}_r)
+ \Bigl[ 
\left( 2\xi -\tfrac{1}{2} \right) \! {\tilde q}_r 
-\xi \bigl(\cosh(r\lambda_1) + 1\bigr)  
\Bigr] 
F'({\tilde q}_r)
\biggr\} 
\,,
\\
- \Delta \langle T_{\chi}{}^\chi \rangle 
&= 
\tfrac12 H^2\sec(\pi \nu) 
\sum_{r = 1}^\infty
\biggl\{
\left[(4\xi-1) \! \left(\frac{m^2}{H^2} + 2\xi \right) + \xi \right] \! F({\tilde q}_r)
\notag
\\
& \hspace{22ex}
+ \Bigl[ 
\left( 4\xi -\tfrac{1}{2} \right) \! {\tilde q}_r 
+\xi \bigl(\cosh(r \lambda_1) -1 \bigr)  
\Bigr] 
F'({\tilde q}_r)
\notag
\\
& \hspace{22ex}
- 2\xi (1 - {\tilde q}_r) \bigl(\cosh(r \lambda_1) + 1\bigr) 
F''({\tilde q}_r)
\biggr\} 
\,,
\end{align}
\end{subequations}
where 
\begin{align}
{\tilde q}_r &:= 1 - \tfrac12 \sinh^2(Ht)\bigl(\cosh(r \lambda_1) -1 \bigr) 
\, . 
\end{align}
The sums in \eqref{Sinh-Chart_VEV} are exponentially convergent, due to the falloff of $F$ at large negative argument. 
Note that the overall numerical factors $\tfrac12$ in \eqref{Sinh-Chart_VEV} have come from combining the $r<0$ image terms with the $r>0$ image terms, whereas in \eqref{Cosh-Chart_VEV} each summand comes from exactly one image term and the overall numerical factor is~$\tfrac14$.

\subsection{Exp scale factor FRW spacetime\label{sec:expchart-stressenergy}}

In the exp scale factor FRW spacetime, an attempt to induce 
a state from the Euclidean vacuum on $dS_2$ by 
an image sum gives for the Wightman function a sum that is not convergent in absolute value, 
as we noted in Section~\ref{sec:expchart-quotientstate}. 
If we set issues of convergence aside for the moment, 
and proceed with the stress-energy tensor as above, working in the chart 
$(t_0,\chi_0)$~\eqref{eq:expchart-metric-cosmoltime}, 
we find that 
\begin{align}
\mathcal{Z}_r \bigl( \mathsf{x}(t_0,\chi_0), \mathsf{x}(t_0',\chi_0') \bigr)
= 
\cosh\bigl(H(t_0 - t_0')\bigr) - \tfrac12 e^{H(t_0 + t_0')} {(\chi_0-\chi_0' + r \lambda_0)}^2
\end{align}
where $r \in \BbbZ \setminus \{0\}$, 
and the image contributions to the stress-energy tensor are given by 
\begin{subequations}
\label{eq:expchart-stressenergy-sums}
\begin{align}
\Delta\langle T_{t}{}^t \rangle 
&= 
\tfrac12 H^2\sec(\pi \nu) 
\sum_{r = 1}^\infty
\biggl\{
- \xi F(\tilde{\tilde q}_r)
+ \Bigl[ 
\left( 2\xi -\tfrac{1}{2} \right) \! \tilde{\tilde q}_r 
-2\xi 
\Bigr] 
F'(\tilde{\tilde q}_r)
\biggr\} 
\,,
\\
- \Delta \langle T_{\chi}{}^\chi \rangle 
&= 
\tfrac12 H^2\sec(\pi \nu) 
\sum_{r = 1}^\infty
\biggl\{
\left[(4\xi-1) \! \left(\frac{m^2}{H^2} + 2\xi \right) + \xi \right] \! F(\tilde{\tilde q}_r)
\notag
\\
& \hspace{22ex}
+
\left( 4\xi -\tfrac{1}{2} \right) \! \tilde{\tilde q}_r 
F'(\tilde{\tilde q}_r)
- 4\xi (1 - \tilde{\tilde q}_r)
F''(\tilde{\tilde q}_r)
\biggr\} 
\,,
\end{align}
\end{subequations}
where 
\begin{align}
\tilde{\tilde q}_r &:= 1 - \tfrac14 e^{2Ht}{(r\lambda_0)}^2 
\,. 
\end{align}
The sums in \eqref{eq:expchart-stressenergy-sums} 
are however not convergent in absolute value. 
While we leave the potential significance of these sums subject to future work, 
we note that the sums depend on $t$ and $\lambda_0$, 
neither of which has an invariant geometric meaning, 
only through the combination $\lambda_0 e^{Ht}$, which does, 
as we discussed in the detector context in Section~\ref{sec:exp-chart-detector}.

\section{Numerical results\label{sec:numerics}}

In this section we give selected numerical results for the detector's response in cosh and exp scale factor FRW spacetimes. 
The associated plots are collected in the Appendix.

\subsection{Cosh scale factor FRW}

\begin{figure}[h!]
\centering
\subfigure[$\Pi_{dS_2}(\mu,1, 0)$]{%
\includegraphics[width=0.46\textwidth]{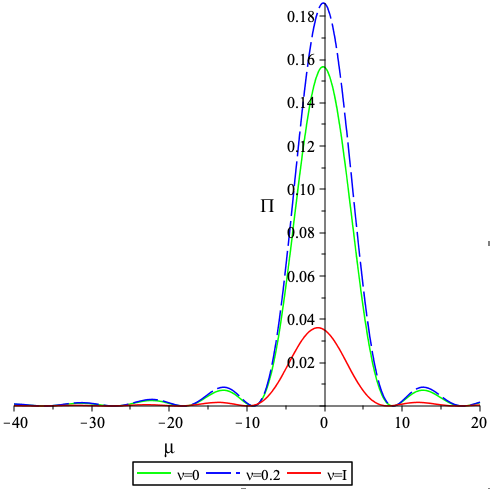} 
\label{Pers:MassiveResFun3D}}
\subfigure[$\Pi_{M_3}(\mu, 1,0)$]{%
\includegraphics[width=0.46\textwidth]{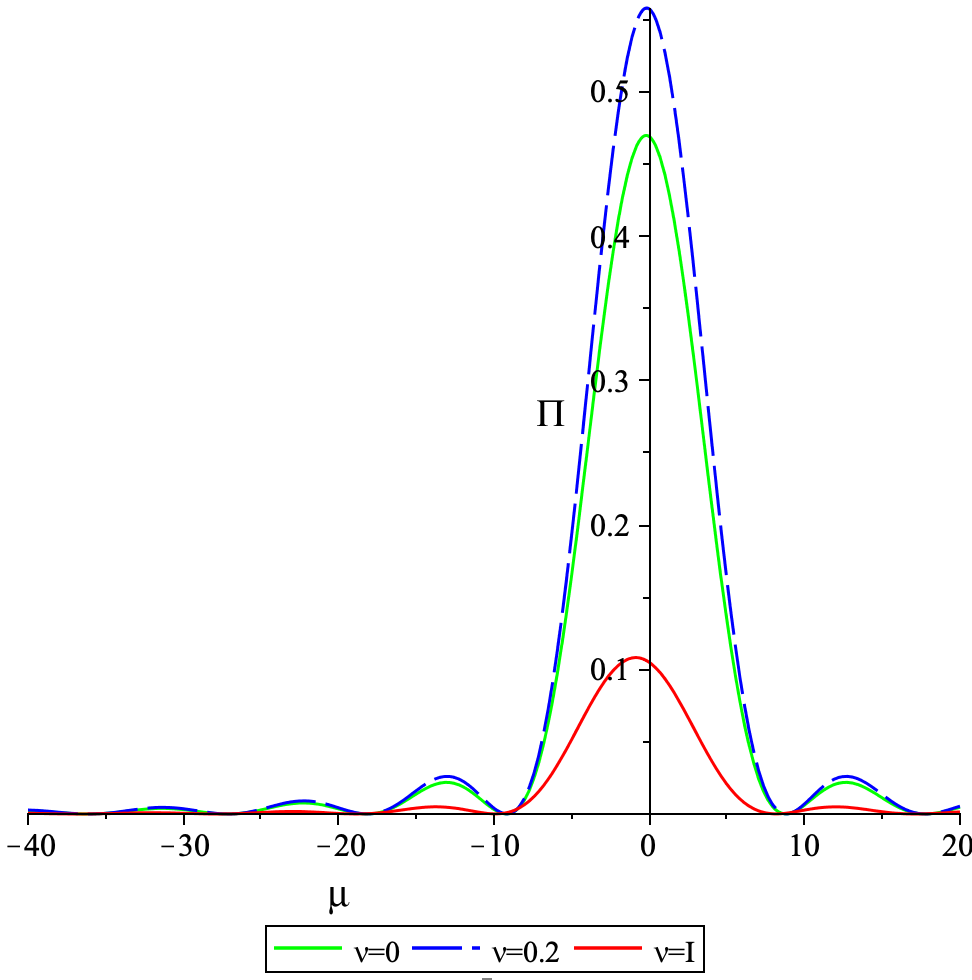} 
\label{CrossSection:MassiveResFun2D}}
\caption{Detector's response in the cosh scale factor FRW spacetime as a function of the energy gap, 
in dimensionless variables, writing $\mathcal{F}(\omega,\tau_1,\tau_0) = H^{-2}\Pi(\omega/H, H\tau_1, H\tau_0)$, and writing $\mu = \omega/H$. 
In part (a) for $\lambda=2\pi$, which is the Euclidean vacuum on~$dS_2$, 
and in part (b) for $\lambda = 2 \pi/3$, which is the induced vacuum on the 
$\BbbZ_3$ quotient~$M_3$, 
for selected values of $\nu$ as shown. 
Evaluated from~\eqref{ResponsFunct-Coshchart-General_Form}.}
\label{Standard_dS_Quotient}
\end{figure}

Figure \ref{Standard_dS_Quotient} shows plots of the detector's response in the cosh scale factor FRW spacetime as a function of the energy gap, for the Euclidean vacuum on~$dS_2$, and for the induced vacuum on the 
$\BbbZ_3$ quotient~$M_3$, for selected values of the parameters as shown. The plots were evaluated numerically from~\eqref{ResponsFunct-Coshchart-General_Form}. 

The high degree of symmetry between excitations and de-excitations indicates that the response in this parameter range is dominated by the switch-on and switch-off effects, 
being significantly different from the long time limit in which the response 
in the Euclidean vacuum satisfies the detailed balance condition in the de~Sitter temperature $H/(2\pi)$~\cite{Gibbons:1977mu}. 

The $\nu$-dependence in the graphs shows that the overall magnitude of the response decreases as the effective mass squared $m^2 + 2\xi H^2$ increases, as one would expect. 
The peak structure in the graphs comes mainly from the spatially constant mode. Similar features were found in \cite{Toussaint:2021czo} for a detector in the spatially compact Milne spacetime.

\subsection{Exp scale factor FRW}

\begin{figure}[h!]
\centering
\subfigure[$\Pi_{osc}^{in}(\mu, 1,0)$ for $\nu=0$ and selected $\lambda_0$.]{%
\includegraphics[width=0.48\textwidth]{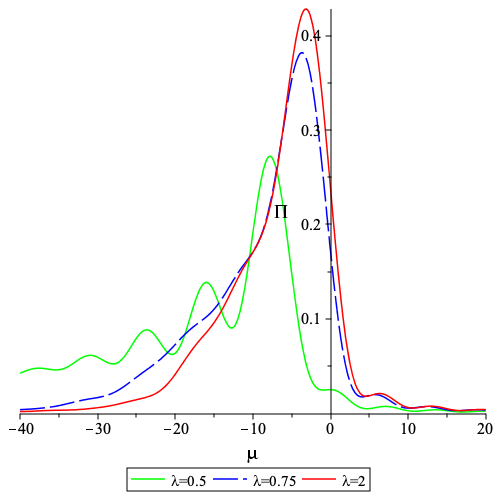} 
\label{CM_OSC_RF_inVac_1a}}
\subfigure[$\Pi_{osc}^{in}(\mu, 2,0)$ for $\nu=0$ and selected $\lambda_0$.]{%
\includegraphics[width=0.48\textwidth]{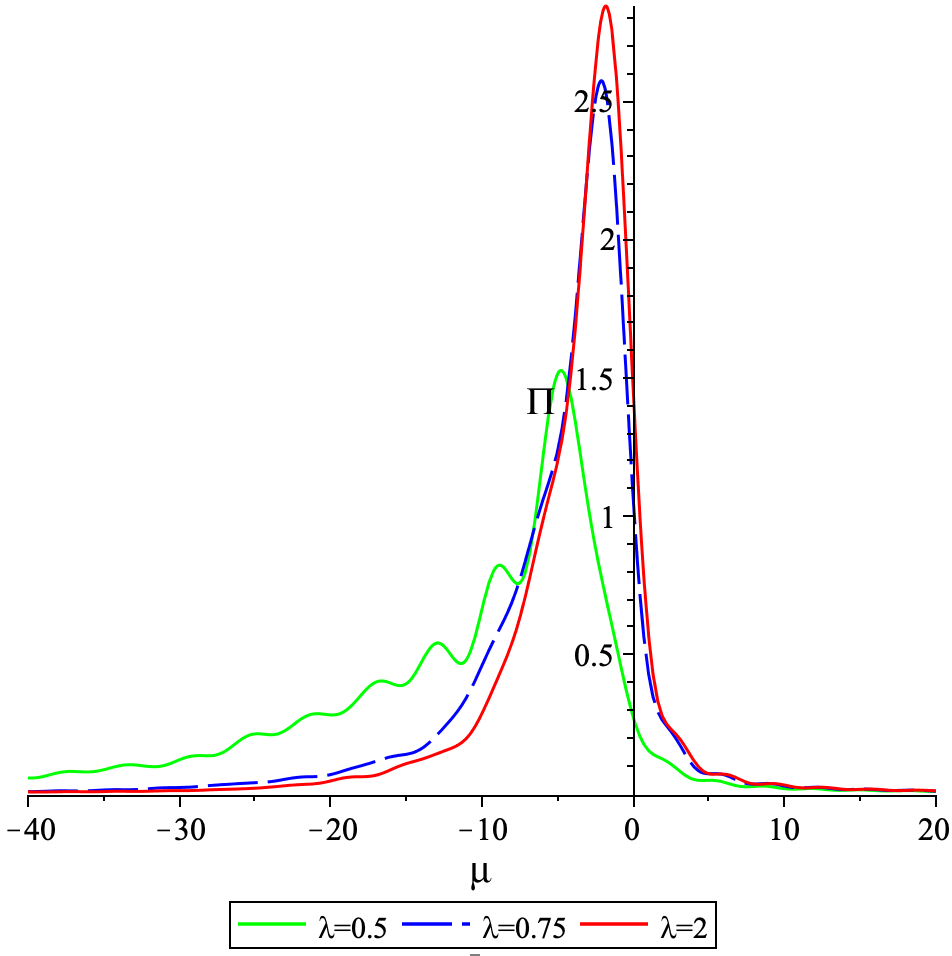} 
\label{CM_OSC_RF_inVac_1b}}
$\phantom{xxx}$\\[3ex]
\subfigure[$\Pi_{osc}^{in}(\mu, 1,0)$ for $\nu=0.3$ and selected $\lambda_0$.]{%
\includegraphics[width=0.48\textwidth]{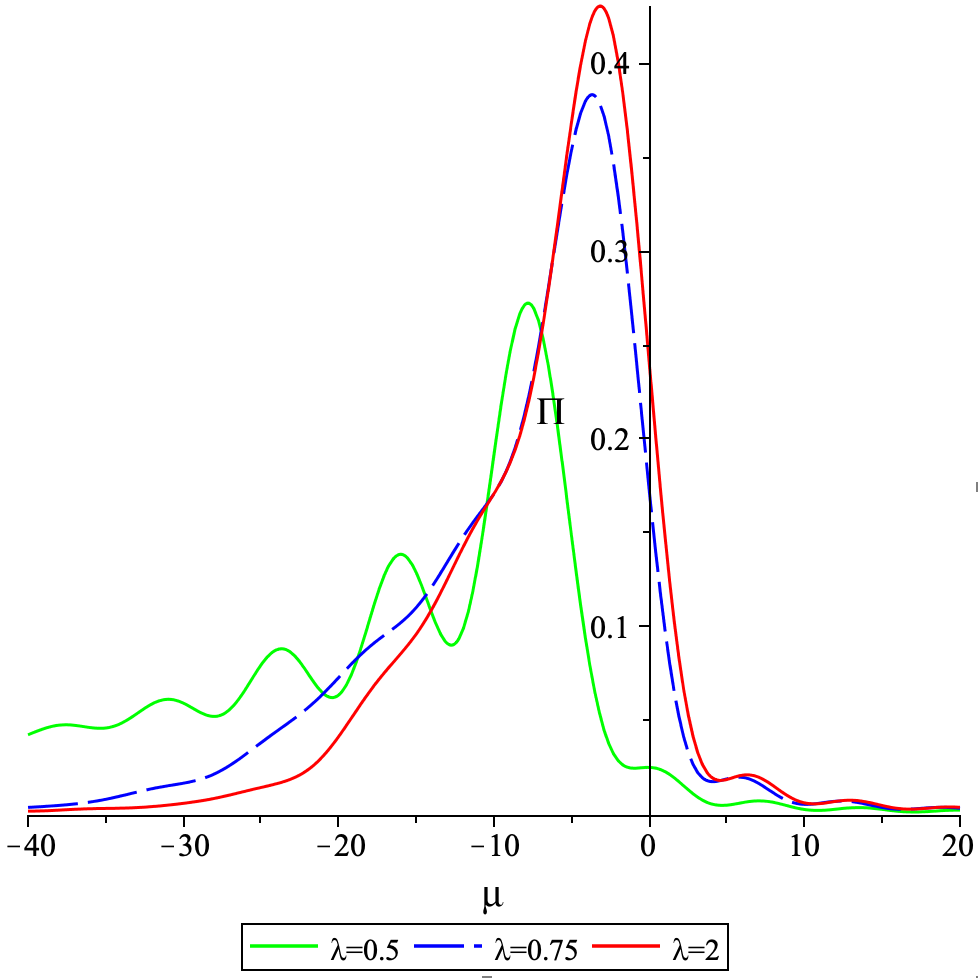} 
\label{CM_OSC_RF_inVac_1c}}
\subfigure[$\Pi_{osc}^{in}(\mu, 1,0)$ for $\nu=0.45$ and selected $\lambda_0$.]{%
\includegraphics[width=0.48\textwidth]{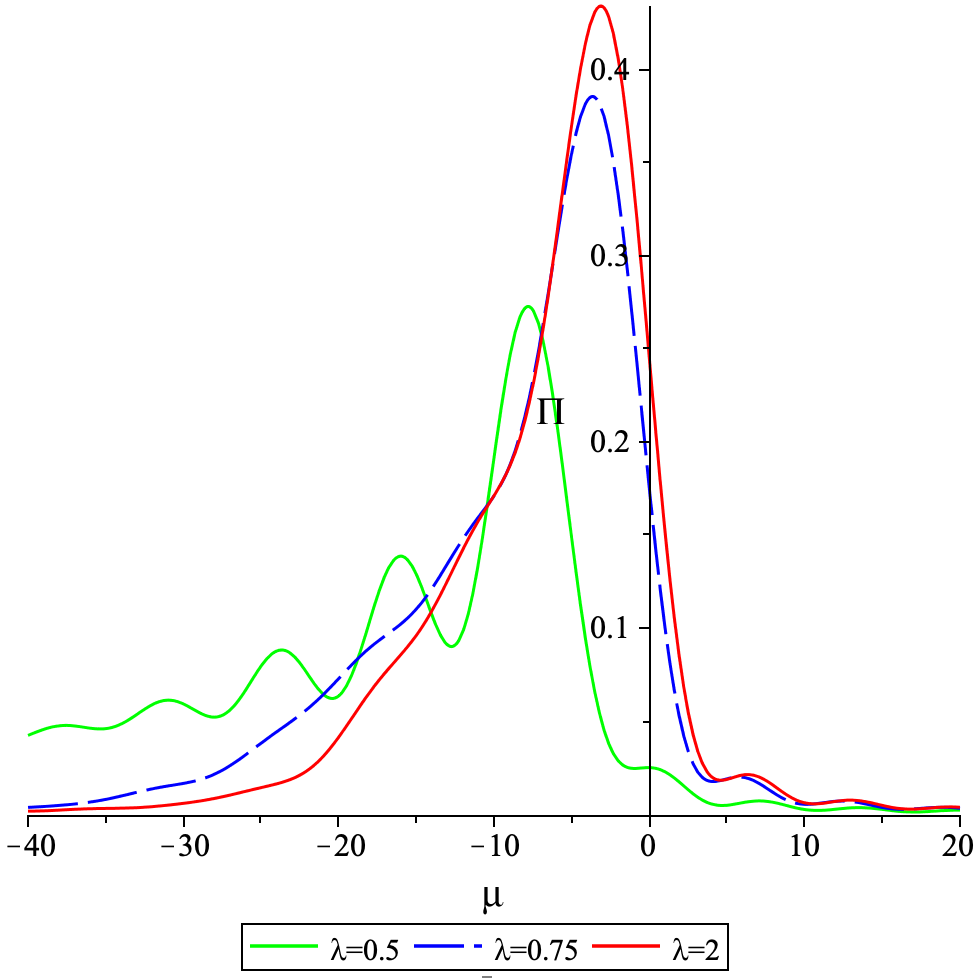} 
\label{CM_OSC_RF_inVac_1d}}
\caption{Spatially nonconstant mode contribution to the detector's response in the exp scale factor FRW spacetime in the ``in'' vacuum, as a function of the energy gap, in the dimensionless variables explained in Figure \ref{Standard_dS_Quotient} caption, with parameter values as shown. 
The legends denote $\lambda_0$ by~$\lambda$.}
\label{CM_OSC_RF}
\end{figure}

Figures \ref{CM_OSC_RF_inVac_1a} and \ref{CM_OSC_RF_inVac_1b} show plots of the detector's response in the exp scale factor FRW spacetime as a function for the energy gap, for the ``in'' vacuum, including only the spatially nonconstant modes, setting $\nu=0$, and using two different values of the duration, for selected values of the circumference parameter~$\lambda_0$. 
The plots show that reducing the value of the circumference parameter shifts the de-excitation peaks towards higher (that is, more negative) de-excitation gaps, and this shift is stronger when the interaction duration is shorter. 
An alternative way to say this, given that the circumference parameter does not have an invariant magnitude, 
is that a detector operating at earlier times has de-excitation peaks at higher (that is, more negative) values of the de-excitation gaps, and this effect is stronger when the interaction duration is shorter. 
Figures \ref{CM_OSC_RF_inVac_1c} and \ref{CM_OSC_RF_inVac_1d} show a similar shift of the de-excitation peaks for selected positive values of~$\nu$. 

\begin{figure}[h!]
\centering
\subfigure[$\Pi_{0}(\mu, 1,0)$ for $\lambda_0=1/2$ and selected $\nu$.]{%
\includegraphics[width=0.48\textwidth]{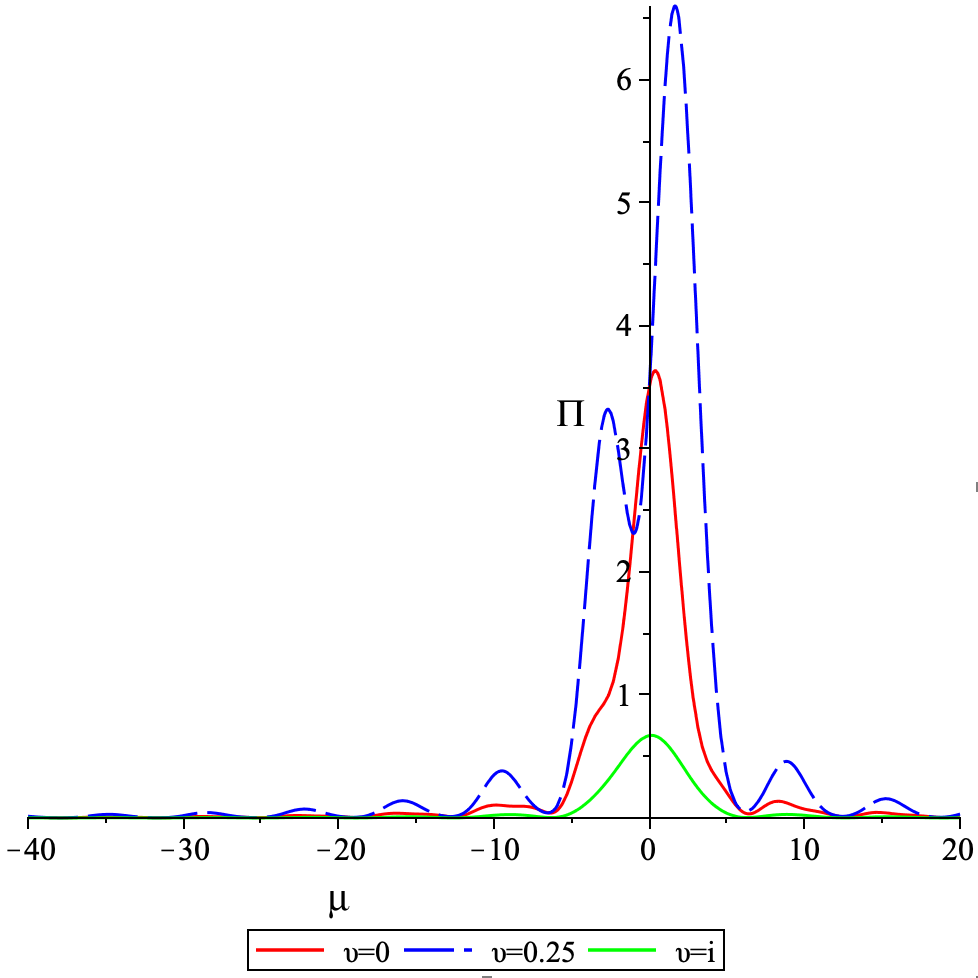} 
\label{CM_ZM_RF_dSVac_1a}}
\subfigure[$\Pi_{0}(\mu, 2,0)$ for $\lambda_0=1/2$ and selected $\nu$.]{%
\includegraphics[width=0.48\textwidth]{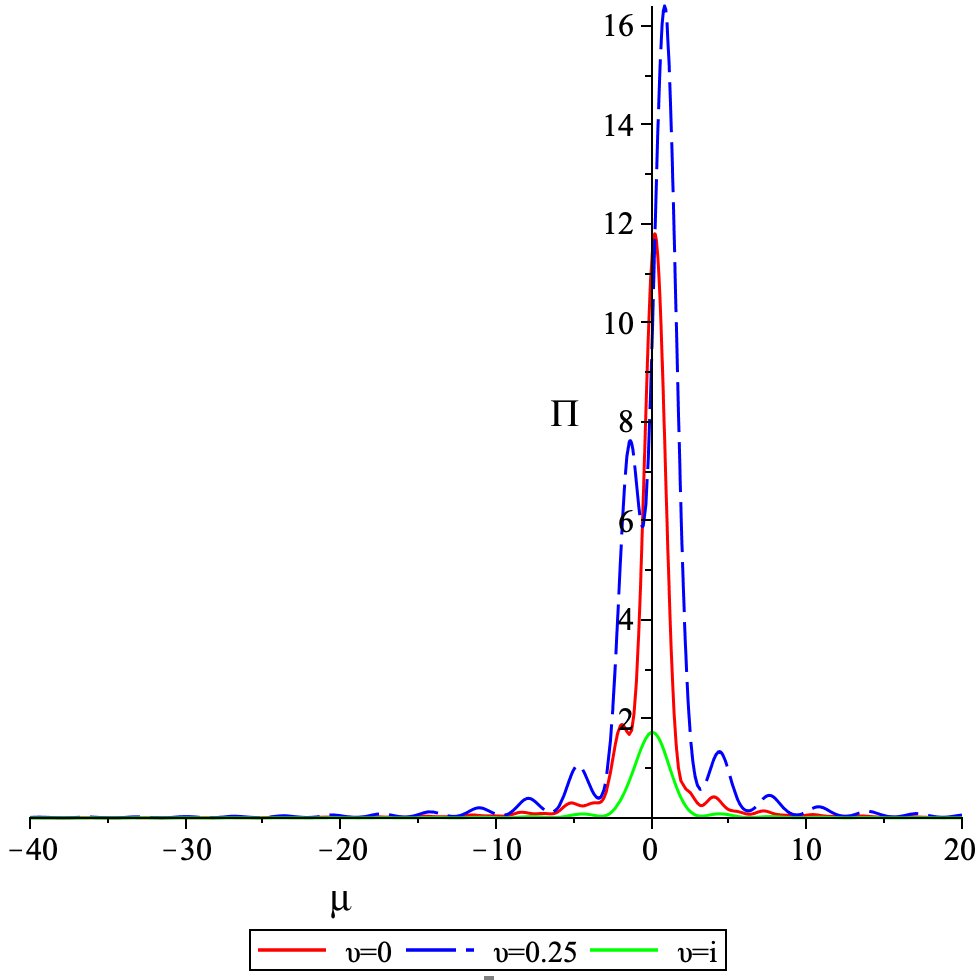} 
\label{CM_ZM_RF_dSVac_1b}}
$\phantom{xxx}$\\[3ex]
\subfigure[$\Pi_{\text{full}}(\mu, 1,0)$ for $\nu=0$ and selected $\lambda_0$.]{%
\includegraphics[width=0.48\textwidth]{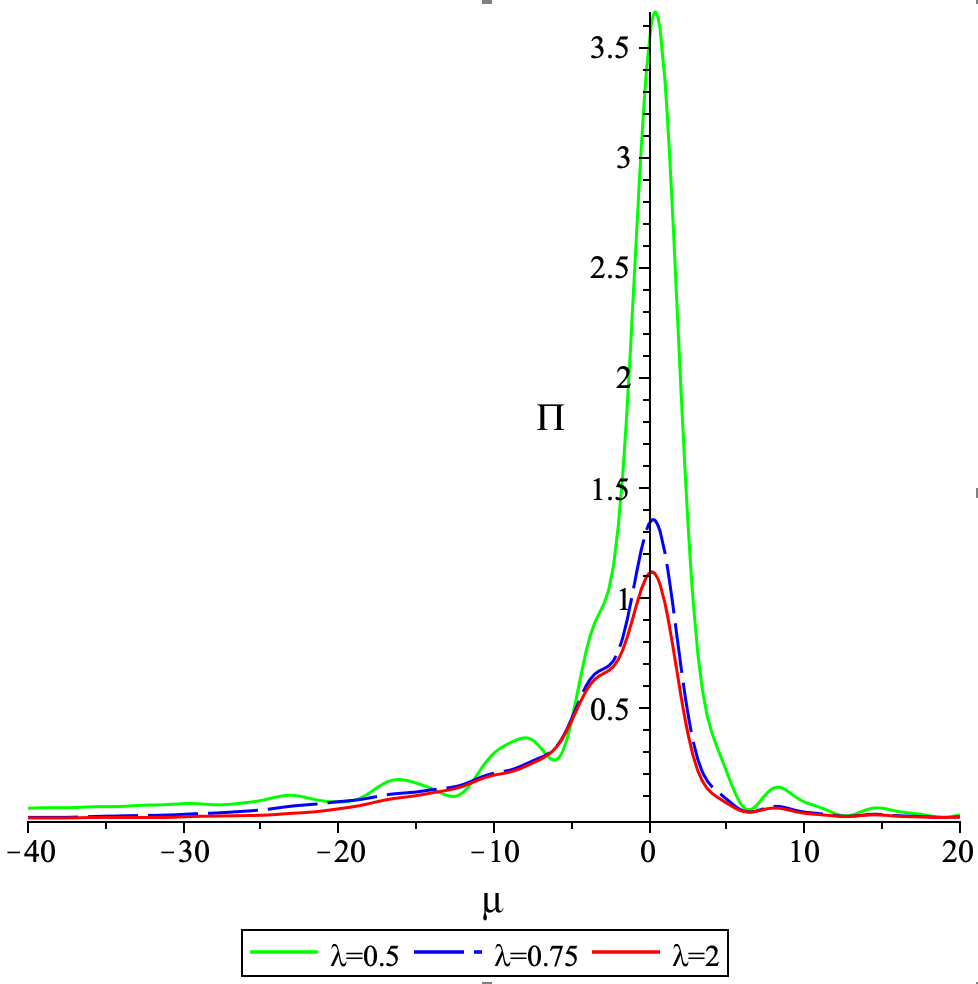} 
\label{CM_Full_Field_RF_dSVac_1c}}
\subfigure[$\Pi_{\text{full}}(\mu, 1,0)$ for $\nu=1/3$ and selected $\lambda_0$.]{%
\includegraphics[width=0.48\textwidth]{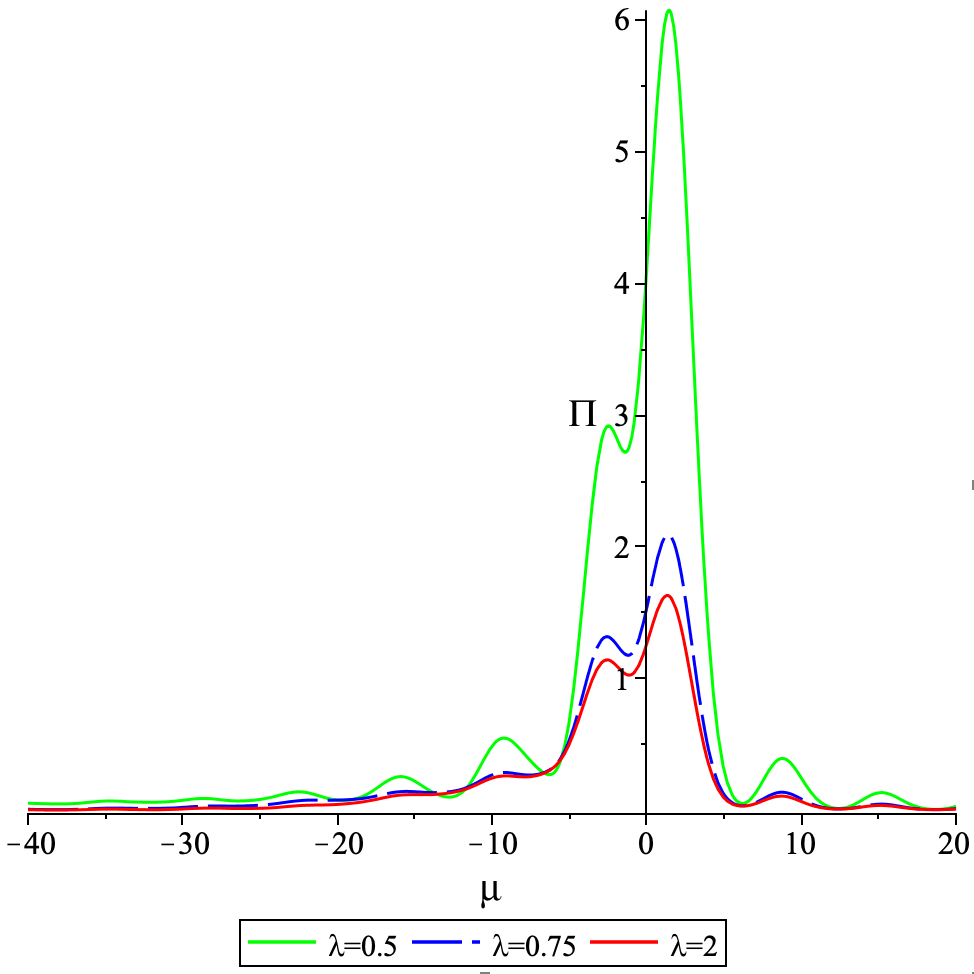} 
\label{CM_Full_Field_RF_dSVac_1d}}
\caption{As in Figure 2, but in parts (a) and (b) for the spatially constant mode contribution, 
and in parts (c) and (d) for the sum of the spatially constant and nonconstant mode contributions.}
\label{CM_ZM_FullField_RF}
\end{figure}

Figures \ref{CM_ZM_RF_dSVac_1a} and \ref{CM_ZM_RF_dSVac_1b} show plots of the contribution to the detector's response from the spatially constant mode only, for fixed circumference parameter but varying~$\nu$, in the ``in'' state. 
The response shows peaks at both excitations and de-excitations, non-symmetrically for low effective mass squared but more symmetrically for high effective mass squared. This indicates that our choice for the spatially constant mode ``in'' state is not close to a no-particle as seen by a local detector. We have also verified that this structure seems not to depend strongly on the size parameter. 

Figures \ref{CM_Full_Field_RF_dSVac_1c} and \ref{CM_Full_Field_RF_dSVac_1d} show plots of the total response, combining the contributions from the spatially constant and nonconstant modes, for selected values of the parameters. 
The spatially constant mode contribution dominates, with minor modulations from the spatially nonconstant modes. 
This is further indication that our choice for the spatially constant mode ``in'' state contains significant structure.

\section{Conclusions\label{sec:conclusions}} 

We have addressed the choice of a quantum state of a real massive scalar field in locally de~Sitter $(1+1)$-dimensional FRW cosmologies with compact spatial sections, with both hyperbolic cosine, hyperbolic sine and exponential dependence of the scale factor on the cosmological time. We employed adiabatic criteria at early times, adiabatic criteria at large momenta, and induction from the Euclidean vacuum on standard global de~Sitter, each of these where applicable. In particular, we found that the early time adiabatic criterion fails for the spatially constant field mode with the sinh and exp scale factors, due to the phenomenon known as a massive zero mode, 
and we discussed a possible alternative criterion for this mode using the late time dependence. 

We found the response function of a cosmologically comoving UDW detector, 
showing that local quantum observations do establish that the detector is not in the standard Euclidean vacuum in standard 
de~Sitter space, and we presented selected numerical results. 
For states induced from the Euclidean vacuum by a quotient construction, we also evaluated the field's stress-energy tensor in terms of image sums, amenable to future analyses by both analytic asymptotic methods and by numerical methods. 

Inducing a state on the FRW spacetime from the Euclidean vacuum on standard de~Sitter spacetime was trivially successful for the cosh scale factor spacetimes that are finite $\BbbZ_p$ quotients of de~Sitter, with $p=2,3,\ldots$, 
and less trivially for the sinh scale factor spacetimes, where the quotienting group is $\BbbZ$ and operates on only a subset of de~Sitter; in the latter case, the infinite image sums for the Wightman function and for the stress-energy were manifestly convergent. A~similar induction attempt for the exp scale factor spacetimes, where the quotienting group is $\BbbZ$ and operates on a different subset of de~Sitter, however produced for the Wightman function and for the stress-energy tensor image sums that are not convergent in absolute value. 
Is this lack of convergence due to some minor technical ambiguity that can be sidestepped by, say, considering derivatives of the field rather than the field itself 
\cite{Raine:1991kc,Raval:1995mb,Lin:2006jw,Wang:2013lex,Juarez-Aubry:2014jba,Bunney:2023vyj}, 
or perhaps a technical artefact that can be cured by an appropriate resummation~\cite{Levi:2015eea}, 
or might it signify some deeper pathology in the sought-for state? 

Our choice to work in $1+1$ dimensions brought in two simplifications. 
First, a geometric simplification was that as the spatial sections are one-dimensional, they have no intrinsic curvature, and 
the compact spatial sections have topology $S^1$ for each of the three expansion laws; further, 
the cosh and sinh scale factor spacetimes are uniquely classified by a spatial circumference parameter that takes arbitrary positive values, whereas the spatial circumference in the exp scale factor spacetime is not associated with an invariant magnitude. Second, a simplification in the detector's response was that we could choose the coupling strength to be constant over the duration of the interaction, with a sudden switch-on and switch-off, without encountering infinities in the detector's transition probabilities, in our first-order perturbation theory treatment. 

A generalisation to spacetime dimension $d+1$ with $d\ge2$, and in particular to our home spacetime dimension $d=3$, 
would need to address a more complicated structure with both of these features. 

First, for $d\ge2$, the spatial sections in the cosh scale factor, sinh scale factor and exp scale factor foliations of standard de Sitter have now respectively positive, negative and zero curvature, with the respective spatial geometries of $S^d$ with the round metric, $\BbbR^d$ with the hyperbolic metric, and $\BbbR^d$ with the Euclidean metric. 
For the sinh and exp scale factors, making the spatial sections compact involves again quotients with continuous parameters, but with a larger choice of spatial topologies; for the cosh scale factor, by contrast, nonstandard spatially compact sections come from quotients of~$S^d$, and these quotients have no continuous parameters. Also, while the quotients are by construction locally spatially homogeneous, not all of them are globally spatially homogeneous. 
A~comprehensive discussion of the $d=3$ case is given in~\cite{wolf-curvature}. 

Second, in the detector's response, a coupling strength that remains constant over the interaction duration produces a finite response function for $d=1$ and $d=2$ but a divergent response function for $d\ge3$, due to the interplay of the sharp switch-on and switch-off against the Wightman function's short distance singularity; see \cite{Hodgkinson:2011pc} for a detailed discussion in the Minkowski vacuum. For $d\ge3$, one would hence need to consider a less abrupt switching, such as the smooth finite duration switchings considered in \cite{Fewster:2015hga,Fewster:2016ewy}, or a $C^n$ switching with some non-negative $n$ \cite{Cong:2020crf,Ng:2021enc}. 
Such switchings however create a new interpretational challenge: how to separate in the detector's response the effects due to the switching profile, which is freely adjustable, from the effects due to the state of the quantum field and the detector's motion, which is the issue of interest? 
An alternative option for $d\ge3$, at least for $d\le5$~\cite{Hodgkinson:2011pc}, 
may be to consider not the transition probability but a transition rate, from which the (divergently) large contribution due to the abrupt switching drops out in a controlled way \cite{Hodgkinson:2011pc,Schlicht:2003iy,Louko:2006zv,Satz:2006kb,Obadia:2007qf,Louko:2007mu}. 

A generalisation to spacetime dimension $d+1$ with $d\ge2$, and in particular to $d=3$, appears hence feasible, but will face more choices in the spacetimes to be considered, and more choices in how the detector's coupling depends on time. 
We leave this generalisation a subject for future work.

\section*{Acknowledgments}

We thank University of Nottingham Ningbo China colleagues Daniele Garrisi and Richard Rankin for interesting discussions
and the anonymous referees for help with the presentational focus.  
The work of JL was supported by United Kingdom Research and Innovation Science and Technology Facilities Council 
[grant number ST/S002227/1]. 
For the purpose of open access, the authors have applied
a CC BY public copyright licence to any Author Accepted Manuscript version arising.

\appendix 

\section*{Appendix: Figures for Section~\ref{sec:numerics}}

In this appendix we collect the figures that are discussed in Section~\ref{sec:numerics}.

$\phantom{xxx}$ 

\newpage

$\phantom{xxx}$ 

\newpage

$\phantom{xxx}$ 

\newpage 

$\phantom{xxx}$

\end{document}